%% file: main.tex
\documentclass[sigconf]{acmart}
\usepackage{soul}
\usepackage{makecell}
\usepackage{subcaption}
\usepackage{xcolor}
\usepackage{multirow}
\usepackage{enumitem}

\newcommand{\revision}[1]{{\color{black}#1}}
\newcommand{\RNum}[1]{\uppercase\expandafter{\romannumeral #1\relax}}
\AtBeginDocument{%
  \providecommand\BibTeX{{%
    \normalfont B\kern-0.5em{\scshape i\kern-0.25em b}\kern-0.8em\TeX}}}

\copyrightyear{2024}
\acmYear{2024}
\setcopyright{acmlicensed}\acmConference[CHI '24]{Proceedings of the CHI Conference on Human Factors in Computing Systems}{May 11--16, 2024}{Honolulu, HI, USA}
\acmBooktitle{Proceedings of the CHI Conference on Human Factors in Computing Systems (CHI '24), May 11--16, 2024, Honolulu, HI, USA}
\acmDOI{10.1145/3613904.3642410}
\acmISBN{979-8-4007-0330-0/24/05}

\begin{document}

\title{Metamorpheus: Interactive, Affective, and Creative Dream Narration Through Metaphorical Visual Storytelling}


\author{Qian Wan}
\affiliation{%
  \institution{City University of Hong Kong}
  \city{Hong Kong}
  \country{China}
}
\email{qianwan3-c@my.cityu.edu.hk}

\author{Xin Feng}
\affiliation{%
  \department{Independent Researcher}
  \city{San Mateo}
  \state{California}
  \country{USA}
}
\email{xinfeng.manx@gmail.com }

\author{Yining Bei}
\affiliation{%
  \department{Department of Architecture}
  \institution{Massachusetts Institute of Technology}
  \city{Cambridge}
  \state{Massachusetts}
  \country{USA}
}
\email{yining25@mit.edu}

\author{Zhiqi Gao}
\affiliation{%
  \institution{Nankai University}
  \city{Tianjin}
  \country{China}
}
\email{gaozhiqi@mail.nankai.edu.cn}

\author{Zhicong Lu}
\affiliation{%
  \institution{City University of Hong Kong}
  \city{Hong Kong}
  \country{China}
}
\email{zhicong.lu@cityu.edu.hk}

\renewcommand{\shortauthors}{Wan et al.}

\input{abstract}



\begin{teaserfigure}
    \centering
    \includegraphics[width=\textwidth]{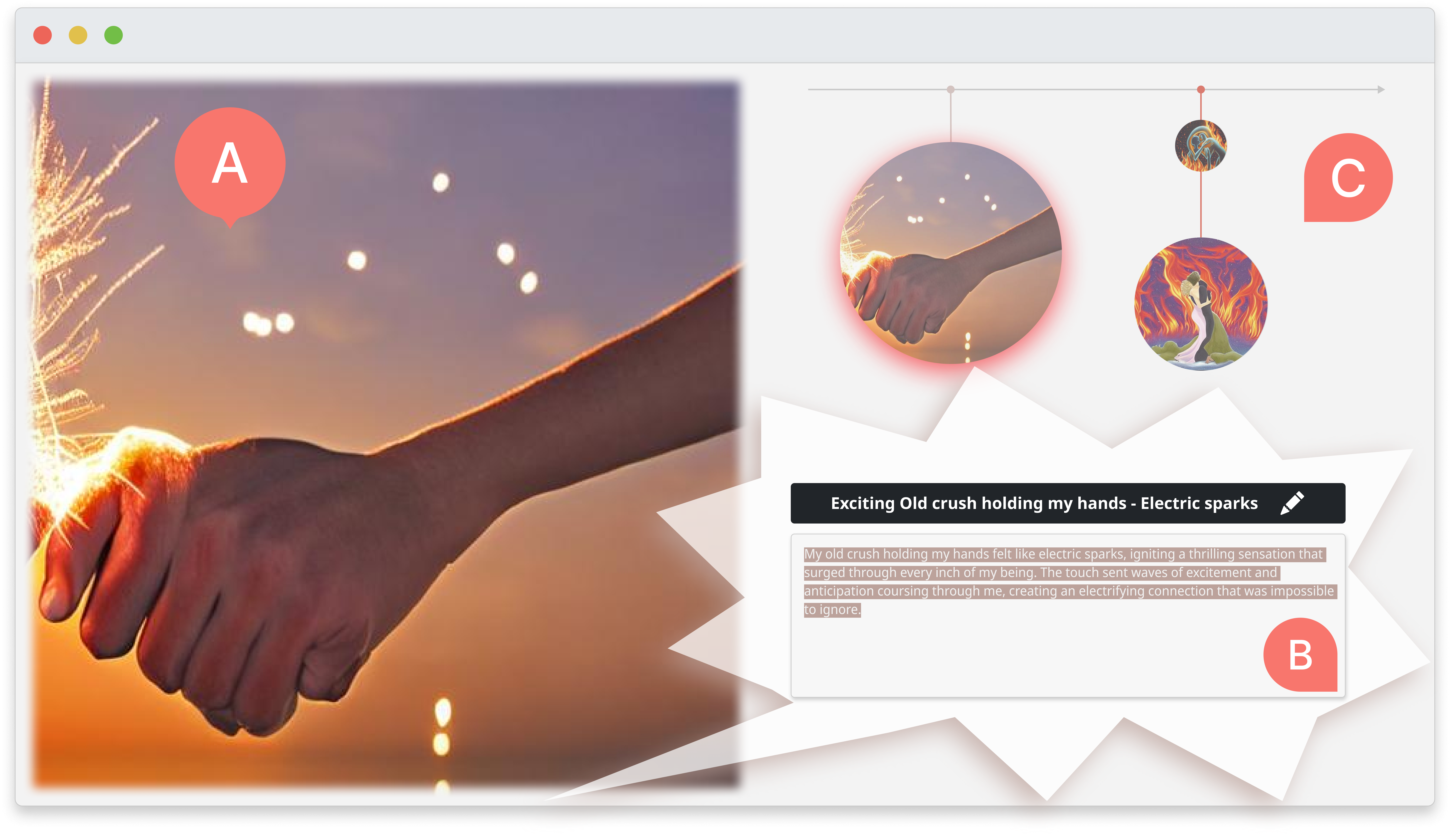}
    \caption{The interface of \textit{Metamorpheus} comprises A) an image in display, B) a text bubble, and C) a storyline visualisation}
    \label{fig:teaser}
\end{teaserfigure}
\maketitle

\input{01-introduction}
\input{02-related_work}
\input{03-metamorpheus}
\input{04-evaluation}
\input{05-findings}
\input{06-discussion}
\input{07-conclusion}


\bibliographystyle{ACM-Reference-Format}
\bibliography{references}


\end{document}

%% file: abstract.tex
\begin{abstract}
Human emotions are essentially molded by lived experiences, from which we construct personalised meaning. The engagement in such meaning-making process has been practiced as an intervention in various psychotherapies to promote wellness.
\revision{Nevertheless, to support recollecting and recounting lived experiences in everyday life remains under explored in HCI. It also remains unknown how technologies such as generative AI models can facilitate the meaning making process, and ultimately support affective mindfulness.} In this paper we present \textit{Metamorpheus}, an affective interface that engages users in a creative visual storytelling of emotional experiences during dreams. \textit{Metamorpheus} arranges the storyline based on a dream's emotional arc, and provokes self-reflection through the creation of metaphorical images and text depictions. The system provides metaphor suggestions, and generates visual metaphors and text depictions using generative AI models, while users can apply generations to recolour and re-arrange the interface to be visually affective. Our experience-centred evaluation manifests that, by interacting with \textit{Metamorpheus}, users can recall their dreams in vivid detail, through which they relive and reflect upon their experiences in a meaningful way.
\end{abstract}

%% file: 01-introduction.tex
\section{Introduction}
The third wave of HCI has generally shifted to a more phenomenological perspective, in which lived experiences of individuals are brought into focus \cite{harrison2007three,prpa2020articulating}. In this trend, HCI designs are becoming increasingly experience-centred \cite{wright2008aesthetics}, and often envisaged or evaluated as experience \cite{mccarthy2004technology}.
Echoing this phenomenological turn, recent advances in affective computing have witnessed a departure from statistical systems that merely detect and record emotions. Instead, researchers are now designing systems situated in our daily routines and social interactions that provide grounds for meaning-making, in order to provoke meaningful self-reflection \cite{rajcic2020mirror}, facilitate communication \cite{murali2021affectivespotlight}, foster social connections \cite{ma2022glancee}, etc.
Nevertheless, considering lived experiences constitute the raw material of human emotions~\cite{ekman1999basic,ekman1992argument,barrett2017theory,gendron2018emotion}, how can computing systems support narration of these affective experiences themselves to facilitate meaning making, for the purpose of self-awareness and self-reflection?

It is known that, putting emotional experiences into words can benefit affective, and mental wellness~\cite{lieberman2007affect,torre2018putting}.
Outside the realm of HCI, expressive therapists have already been using creative and artistic expression of our lived experiences as a treatment for mental health conditions such as nightmares \cite{krakow2006clinical}, trauma \cite{pifalo2007jogging}, depression \cite{maratos2008music}, etc. Through various forms of creative expression, these therapies aim to
``\textit{facilitate clients' discovery of personal meaning}''
, which, as Malchiodi put it \cite{malchiodi2013expressive}, ``\textit{may deepen into greater self-understanding or may be transformed, resulting in emotional reparation, resolution of conflicts, and a sense of well-being.}'' Writing therapists, for instance, might engage clients in a narration of their emotional experiences (e.g., a traumatic experience) by journal or poetry \cite{Adams1999,ostriker2018poetry} to attenuate negative experiences \cite{torre2018putting,pennebaker1997writing,wilson1991thinking}.

Previously, HCI researchers also found that creative activities such as bullet journaling~\cite{tholander2020crafting,ayobi2018flexible} or visual storytelling~\cite{zhang2022storydrawer,halperin2023envisioning,zarei2020investigating} can facilitate self-reflection and promote mindfulness through a creative meaning-making process. To track emotions, for example, bullet journalists were found to come up with individualised, decorative, and artistic representations of moods, such as Mandala, Origami, text narrative, etc \cite{ayobi2018flexible}.
Compared to emotion recognition algorithms~\cite{andalibi2020human}, creative expression as a way of emotion tracking affords a sense of agency and ownership, where users have control over their personal data and invest personal meaning to their creation ~\cite{ayobi2018flexible,tholander2020crafting}. \revision{However, there still lacks an affective interface that enables common users, even those without any art expertise, to creatively relate an emotional experience for meaningful self-reflection.} We are thus particularly interested in how computer-supported creative narration of affective experiences, in lieu of a therapist, can help users construct personalised meaning, provoke meaningful self-reflection, and ultimately, promote affective mindfulness.

In this paper, we explore the paradigm of visual storytelling to provoke meaningful self-reflection through creation, using dreams as the material. Visual storytelling is commonly used in HCI to promote mental well-being (e.g., \cite{zhang2022storydrawer,zarei2020investigating}) that combines two common modalities: text and visuals. Dreams are believed to be enactment and dramatization of waking-life experiences \cite{domhoff2017invasion,bell2011personality} that provides sources for self-discovery and self-understanding \cite{de1994ballantine,hill1997dream}. They are also theorized to be influenced by waking emotional concerns, and pivotal in emotional wellness, which makes them valuable materials to promote affective mindfulness.

To support creative expression of emotions, we present \textit{Metamorpheus}, an affective interface that engages users in the creation of metaphorical visual stories of dreams. The system arranges a visual story based on users' recollection of their emotions in dream narratives, and facilitates self-discovery and self-reflection by enabling users to co-create metaphorical text depictions and images of visual metaphors of their emotions with generative AI models. \textit{Metamorpheus} also allows users to alter the interface by relocating and resizing each emotional scene over a storyline visualisation, and applying colour filters extracted from created images based on the colour dominance, with the goal of encouraging them to further construct personalised meaning out of co-created text and images.

We deployed \textit{Metamorpheus} in a controlled lab environment as a technology probe, and conducted a phenomenological evaluation of the interaction experience. The study results shed light on how \textit{Metamorpheus} afforded interactive meaning co-construction through co-creative narration, which promoted mindfulness, created connectedness, and motivated dream-sharing in an innocuous manner. It also informs future design to incorporate human-machine co-creation as a means of meaning co-construction for psychotherapy practice and personal informatics.

Our paper makes the following contributions to HCI:
\begin{itemize}[topsep=0pt]
\item the design and implementation of \textit{Metamorpheus}, an affective interface that engages users in a creative narration of emotional experiences in a dream.
\item a nuanced understanding of how \textit{Metamorpheus} afforded interactive meaning co-construction experience, by which it provokes meaningful self-reflection.
\end{itemize}

\input{visualisation}

%% file: visualisation.tex
\begin{figure}
    \centering
    \includegraphics[width=\linewidth]{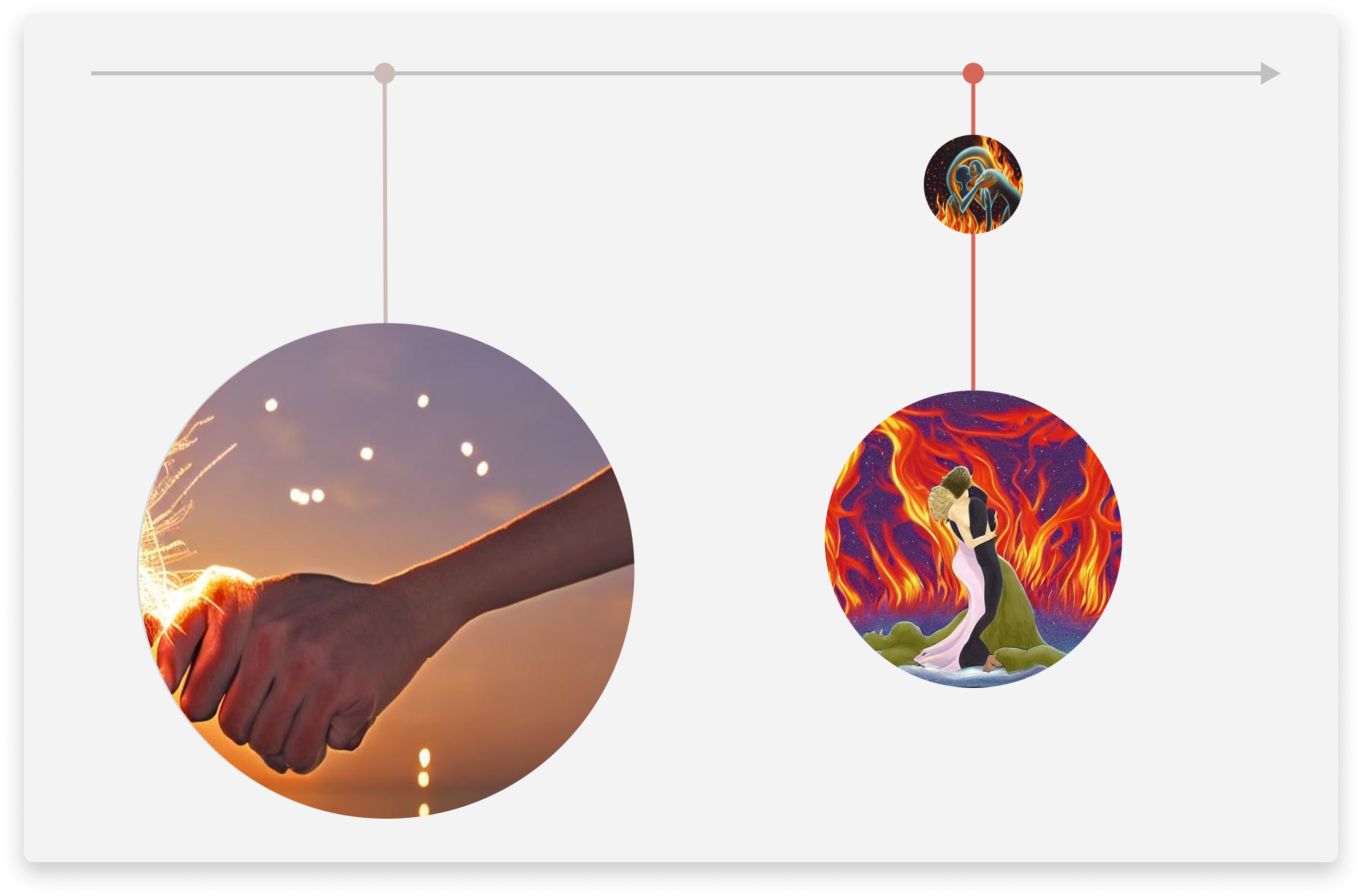}
    \caption{The storyline visualisation on the interface arranges the dream narrative in a linear order. Each metaphorical scene is represented as an anchor point on a horizontal axis. The current generated images dangle from the anchor points, and previous generations are preserved above. Users can click on previous generations to switch it into display.}
    \label{fig:visualisation}
\end{figure}

%% file: 02-related_work.tex
\section{background and related work}
This work was motivated by previous literature on affective computing, creative journaling, and dreams. In this section, we first introduce the background of emotion theories and affective computing, with a focus on recent constructivist models and meaning-oriented frameworks. We then briefly review previous work of creative journaling and dreams in HCI to explain our design rationale.

\subsection{Emotions and Meaning Making}
The broad arc of affective computing has generally shifted away from essentialist views of emotions, to a more constructivist stance. When the concept of ``affective computing'' was first introduced by Rosalind W. Picard in 1997, it was formulated as a problem of pattern recognition and synthesis~\cite{picard2000affective}, as influenced by the classical emotion theories that argue for the existence of \textit{basic} categories that are universal, fundamental, but discrete and physiologically distinct~\cite{ekman1999basic,ekman1992argument}.
\revision{However, this essentialist view was challenged by researchers, such as Lisa Feldman Barrett, by emphasizing that emotion categories are shaped by contextual factors~\cite{barrett2006solving,barrett2013large,lindquist2012brain,barrett2007mice}.
Alternatively, constructivist theories have been proposed and formulated in its stead ~\cite{barrett2017theory,barrett2014psychological,russell2003core,russell2015greater}, in which instances of emotion are rejected to be \textit{natural kinds}, but rather constructed by our brain in the moment by assigning meaning to sensory inputs, as informed by past experiences~\cite{barrett2017emotions,barrett2006solving}.}

The constructivist shift tacitly coincided with the emergence of the third-wave HCI research, initially conceptualised as the phenomenologically situated paradigm, which is ``\textit{a focus on meaning and meaning creation}'', ``\textit{based on human experience}'', and ``\textit{therefore represented through multiple perspectives, and the relationship amongst those perspectives}'' \cite{harrison2007three}. Following these trends, Boehner et al. challenged the traditional informational model of emotions that presupposes the existence of discrete emotions, and offered an alternative account that models emotions as a social and cultural product experienced through interactions~\cite{boehner2005affect}. They argued that such model leads to new goals for affective computing, wherein systems should support users in making meaning of their own emotions instead of merely sensing and transmitting prescribed emotion labels.

Inspired by the interactional model, recent designs of affective interfaces are increasingly socially situated, experience-centred, and meaning-oriented \cite{rajcic2020mirror,murali2021affectivespotlight,ma2022glancee}. For example, Rajcic et al. \cite{rajcic2020mirror,rajcic2020mirror_meaning} designed a smart mirror that engages users in making meaning of their emotional states to promote awareness and reflection. The mirror performs an emotion recognition algorithm and composes a poetry using a large language model (GPT-2) based on the detected emotion states to foster meaningful experience. Their user study used Mekler and Hornb{\ae}k's~\cite{mekler2019framework} framework to evaluate the experience of meaning, which comprises five components: \textit{Connectedness}, \textit{Purpose}, \textit{Coherence}, \textit{Resonance}, and \textit{Significance}. It revealed that the creative and open-ended nature of poetry allowed users to develop personalised meaning from the mirror's generations.

The design of our system, \textit{Metamorpheus}, is motivated by the recent constructivist conceptualisations of emotions and the phenomenological paradigm. Drawing upon insights of Boehner et al.~\cite{boehner2005affect}, our design hones in on the lived experience, the raw material of human emotions, and aims to provoke meaningful reflection through creative narration. Because meaning as a quality of interaction is often elusive, we also used Mekler and Hornb{\ae}k's framework in our phenomenological evaluation, which provides an constructivist understanding of how users constructed meaning from both the interaction experience and end results.

\subsection{Creative Journaling of Personal Information}
Throughout the history of mankind, creative and artistic activities have consistently been associated with healing and therapies~\cite{mcniff1981arts,mcniff1992art}. Historical evidence shows that the practice of arts as therapies dates back to thousands of years ago, when Egyptians encouraged those with mental illness to engage with artistic expression~\cite{fleshman1981arts}, and Greeks used drama and music as a reparative form of treatment~\cite{gladding1992counseling}. Roughly during the 1930s and 1940s, expressive therapies gained wider recognition, as psychotherapists began to realise that self-expression through non-verbal channels (e.g., painting, music, dance, etc.) might benefit people with severe mental illness~\cite{malchiodi2013expressive}. To this day, scientific evidence continues to surface supporting the effectiveness of these therapeutic practices of artistic expression~\cite{slayton2010outcome,stuckey2010connection,reynolds2000effectiveness}. Various theories have also been put forward in an attempt to explain the exact mechanism through which expressive therapies improve mental well-being. For affective wellness in particular, it is believed that artistic expression can translate into emotional reparation, awareness, and resilience~\cite{malchiodi2013expressive,pearson2008using}, by supporting and facilitating clients' personal meaning-making process~\cite{boals2012use}.

Previously, HCI researchers have also found similar effects in creative journaling of personal information~\cite{tholander2020crafting,ayobi2018flexible,ayobi2020trackly}. Early work by Petrelli et al.~\cite{petrelli2009making} reveals that people are more interested in creatively reconstructing memories from carefully selected cues than exhaustively digitally recording them. Therefore, they argued that future autobiographical technologies should aim to support \textit{active selection}, \textit{creativity}, and \textit{meaning building}. Later studies of bullet journals by Ayobi et al.~\cite{ayobi2018flexible} further suggests that bullet journalists engage in mindful reflective thinking through the design of personally meaningful textual, numeric, and symbolic representations of different types of trackers. In particular, they reported that creative crafting of individualised, and visually-appealing mood trackers can have therapeutic effects on their emotional well-being.

It is argued that, as a way of managing personal information, the creative nature of bullet journaling affords a sense of agency that users feel like having full control of their data. Therefore people often take ownership of their creation~\cite{ayobi2018flexible,tholander2020crafting}, and invest personalised meaning into them in an reflective manner ~\cite{tholander2020crafting}. Nevertheless, to our best knowledge, computing systems, especially recent generative AI models have yet not been deployed to interactively participate in the creative journaling of emotional experiences to facilitate meaning-making. The design of AI-powered journaling can engage average users in a creative meaning-making process, but also risks impeding their sense of agency ~\cite{elsden2016s,ayobi2020trackly}. Previous design probes of AI co-creation for art therapies also reveals the concern of powerful generative AI models being ``\textit{overpowering}''~\cite{du2024deepthink}. In the design of \textit{Metamorpheus}, we consider balancing the two by introducing generative AI for meaning making through the creation of metaphors, while keeping the interaction process and the interface creative and open to interpretations. The evaluation of \textit{Metamorpheus} probes into the potential of computer-supported, and particularly AI-supported meaning co-construction of lived experiences for mental wellness.


\subsection{Dream and Dream Journaling}
While the exact origin and function still largely remain a mystery to neuroscience~\cite{givrad2016dream,hoel2021overfitted,palagini2011sleep}, researchers have already come up with diverse methods to study dreams, such as neuroimaging \cite{kusse2010neuroimaging}, content analysis (notably the Hall/Van de Castle coding system~\cite{hall1966content}), and quantitative analysis of coded dream reports \cite{domhoff1996finding} using the Hall/Van de Castle system. A variety of theories have also been put forward to explain the dreaming process, the most notable being the \textit{continuity hypothesis}, which suggests dreams ``\textit{enact and dramatize}'' waking-life experiences~\cite{domhoff2017invasion,bell2011personality}. This means that dreams can serve as a valuable material for self-discovery and self-understanding, as dreams reflect the same concerns of waking thoughts in a more dramatic way~\cite{hoefer2022personal}.
Dream experience is also hypothesized to be entangled with our emotions, and vice versa. Preliminary psychological and neuroimaging findings suggest that, dreams are influenced by dreamers' waking emotional concerns~\cite{cartwright2006relation}, and dreaming might play a pivotal role in the emotional regulation and emotional memory consolidation~\cite{scarpelli2019functional}.

In practice, dreams have already been widely used in therapies~\cite{hill1996working,keller1995use}, especially as a treatment for~\textit{nightmare disorder}~\cite{gieselmann2019aetiology}. In a typical Imagery Rehearsal Therapy (IRT), for example, therapists would ask clients to recount their dreams and rewrite the narrative such that it is no longer a nightmare~\cite{krakow2006clinical,krakow1995imagery}. The revised dream needs to be mentally rehearsed for several minutes a day~\cite{halliday1987direct}, so that it begins to occur during the sleep in place of the nightmare, to improve the overall psychological well-being.

Recently, the significance of dreams has also drawn attention from HCI researchers. Hoefer et al. was argubly the first to study dreams as a form of health data used in a personal informatics system~\cite{hoefer2022personal}. Their survey study presented comprehensive challenges for computers to support dream tracking and dream journaling. For instance, dreams are easily forgotten, hard to capture as an entire lived experience, and often personalised and open to interpretations. Perhaps for these reasons, they concluded that technologies to support dream recall and dream narration were nearly non-existent except for note-taking features.

The design of \textit{Metamorpheus} fills this gap by introducing human-computer co-creation of visual stories as a method for interactive dream narration. Our experience-centred evaluation demonstrates that \textit{Metamorpheus} not only facilitates dream recall, but also provokes meaningful self-reflection.

\input{form}

%% file: form.tex
\begin{figure*}
    \centering
    \includegraphics[width=.95\textwidth]{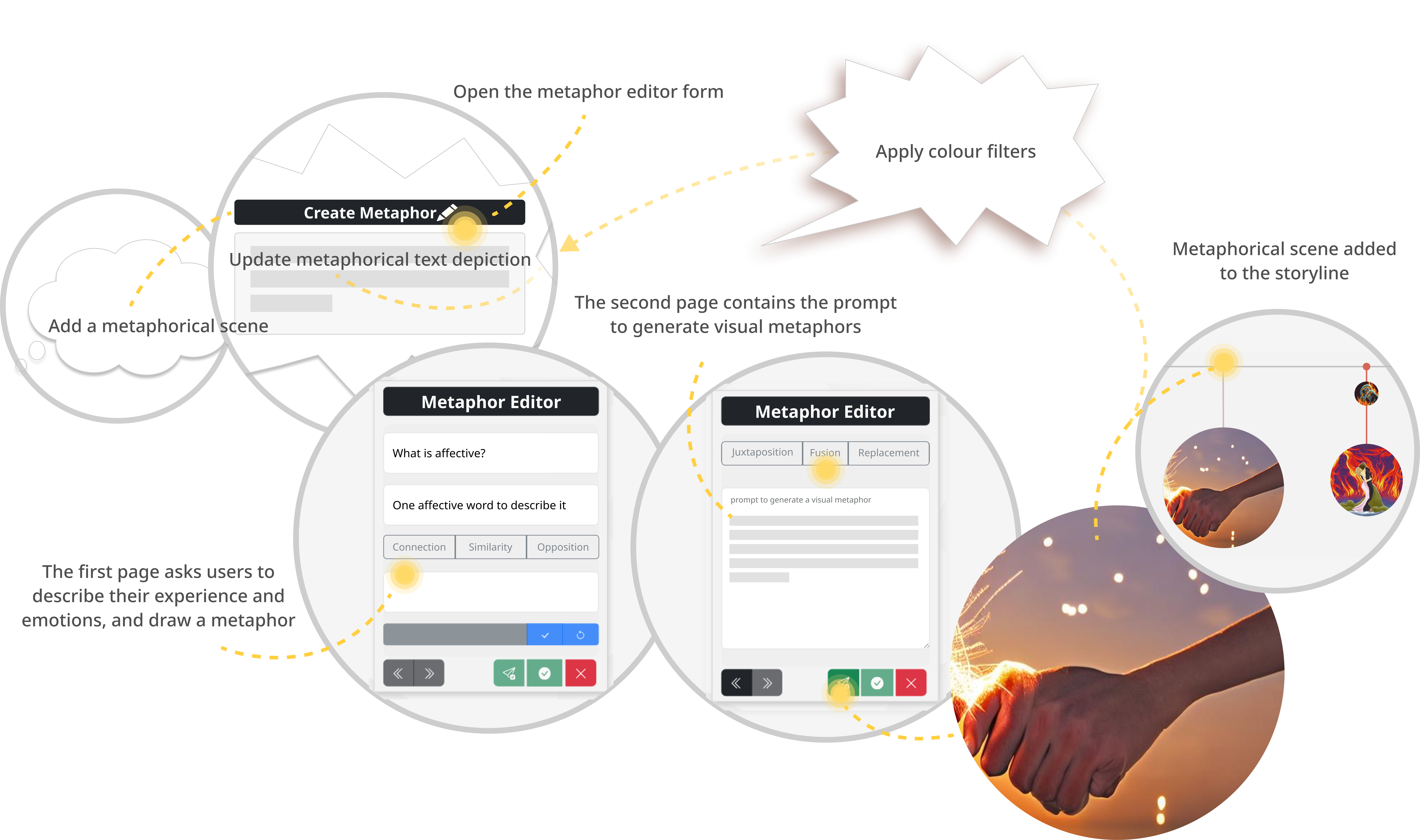}
    \caption{The metaphor editing process.}
    \label{fig:form}
\end{figure*}


%% file: 03-metamorpheus.tex
\input{scenario}

\section{Designing Metamorpheus}
The design of \textit{Metamorpheus} was informed by a review of affective expression in existing visual narratives. The creation of affective metaphors ended up being chosen as the main feature of \textit{Metamorpheus} to develop a visual story. In this section we introduce our design considerations, and how they were implemented in practice.

\subsection{Design Considerations}
The design of \textit{Metamorpheus} attempts to engage users in the creation of a visual story to recount their emotions during a dream. For the purpose of affective reflection, we expect the system to be creative and open to interpretations for self-expression, and meanwhile remains accessible to common users with little art or design expertise. To this end, the authors, three HCI researchers plus two designers (one UX designer \& one architecture designer), began by surveying papers on devices for affective expression in previous visual narratives such as films, cartoons, anime, comics, photography, etc.

A discussion session was later held among authors to review each paper and design approach we surveyed. We concurred to filter out approaches that are: \RNum{1}) too individualised and arbitrary for a visual storytelling system (e.g., pictorial runes in comics \cite{forceville2011pictorial}), \RNum{2}) too demanding that might seem distracting for storytelling (e.g., emoji or meme creation \cite{evans2017emoji,terzimehic2021memeories}), and \RNum{3}) not creative or open-ended enough for the purpose of self-expression and self-reflection (e.g., colours, shapes, filters,  style transfer, etc).

In the end, we opted for the creation of affective metaphors to help develop visual stories in \textit{Metamorpheus}. The creation of metaphors adapts better to affective visual storytelling because it is a cross-modal and emotionally arousing device for artistic expression that has been widely used across various mediums of communication and arts, even including HCI designs \cite{reed2023negotiating}.
In clinical practice, metaphors also serve as an effective way of communicating lived experiences~\cite{stilwell2021painful}.
Additionally, we already have a well-defined design space~\cite{phillips2004beyond} to scaffold the creation process, including both meaning and visual structures of metaphors. We also have corresponding creativity support tools~\cite{kang2021metamap,gero2019metaphoria}, and text-to-image AI models \cite{rombach2022high} readily available to support users.

Nevertheless, adapting a metaphor creation to an affective dream narration process requires additional considerations. Dreams are known to be hard to recall, hard to express, and open to interpretations \cite{hoefer2022personal}. Therefore the creation of metaphors must encourage users to verbalise and visualise their emotions within dream scenes, and engage them in a meaning making process by prompting recollection and reflection. In this sense, \ul{the creation process needs to be anchored in the expression of emotions in a more \textit{articulate} manner across both \textit{semantic} and \textit{visual} modalities} \textbf{(G1)}. Otherwise, the system might seem distracting, unfocused, and even meaningless for dream narration. For this purpose, we can apply ChatGPT suggestions and text-to-image models based on the previous design space~\cite{phillips2004beyond} to help draw metaphors to articulate emotions, but it might risk impeding users' sense of agency or ownership that is pivital in stimulating personalised meaning-making and reflective behaviours. Therefore, \ul{to facilitate the meaning making process, the articulated expression itself has to remain \textit{artistic}, \textit{creative} and \textit{open to interpretations}, throughout the interaction} \textbf{(G2)}.

As mentioned, the two goals might sometimes contradict each other, because articulating dream scenes through a metaphorical device, with any forms of automation or scaffolding, might limit the users' agency and ownership of interpreting dream content. In what follows, we describe how we designed \textit{Metamorpheus} so as to strike a balance between the two goals, and how we deployed the system as a probe into the intersection between the \textit{sense of agency} and \textit{meaningful expression}.

\subsection{User Interface}
\textit{Metamorpheus} divides a complete visual story into multiple literal or metaphorical scenes. Each literal scene contains only text descriptions, while metaphorical scenes store both metaphorical text depictions and generated images of visual metaphors. The editing and presentation of the visual story are integrated into one user interface, which comprises three common UI elements in a visual story: a text bubble, a storyline visualisation, and an image of a visual metaphor in display, as shown in \autoref{fig:teaser}.

\subsubsection{Text Bubble \& Metaphor Editing}
The text bubble on the interface displays text descriptions of each scene. For metaphorical scenes, an extra title is displayed to indicate the metaphor drawn: the original affective concept, and a new metaphorical concept. To make it attentive to emotional changes in narratives for \textbf{G1}, two different shapes were designed to distinguish literal and metaphorical scenes. As informed by \cite{aoki2022emoballoon}, we assumed that spiky text bubbles imply higher emotional arousal than rounded ones. Therefore, we use spiky text bubbles for metaphorical scenes, and rounded bubbles for literal scenes, as shown in \autoref{fig:bubbles}.

For a metaphorical scene, users can click to open the metaphor editor. To keep the editing anchored in the emotion recollection and expression for \textbf{G1}, we require users to specify what is affective, the description of the affective element (one or more adjectives), the metaphor to draw, and prompts to generate an image of the visual metaphor (see \autoref{fig:form}), which is to some extent similar to affect labeling~\cite{torre2018putting}. We also scaffold the creation of visual metaphors by applying the design space in \cite{phillips2004beyond}, where users are allowed to select the type of meaning (connection, similarity, and opposition), and visual structures (juxtaposition, fusion, and replacement) to automatically request suggestions of metaphors from ChatGPT, and generate template prompts for a text-to-image model. After a visual metaphor is accepted, the storyline visualisation will be updated accordingly, and a piece of metaphorical text will be generated by ChatGPT to depict the metaphor drawn and added to the text description of the scene.

\subsubsection{Storyline Visualisation \& Colour Filters}
The storyline visualisation on the interface arranges the emotions of a dream narrative in a linear order, where each metaphorical scene is represented as an anchor point on a horizontal axis. The current generated images of each scene dangle from the anchor point, and previous generations are preserved above. To make the visualisation more creative and open to interpretations for \textbf{G2}, we allow users to relocate or resize each of the dangling images to re-arrange the visualisation and construct their personalised meaning.

The image display area holds the current generated image of a metaphorical scene in focus. For each of the image generated, we extracted 8 dominant colours for users to apply customised ``colour filters'' over the interface to support our \textbf{G2}, as inspired by~\cite{ayobi2020trackly}. The extraction was achieved by a clustering algorithm similar to K-means. For details we refer our readers to ~\cite{chang2015palette}. By default, the most dominant colour will be selected from the image in display to recolour the interface, including the colour of text bubbles' shadow, anchor points and dangling lines of scenes, and background of generated text description of metaphors.  Users can click on the image in display to open a colour picker where they can select different colours to update the filters.

\subsection{Example Scenario}
We provide an example to illustrate the creation workflow of \textit{Metamorpheus} using a dream narrative sampled from Dream Bank (an online collection of over 22000 dream reports) \footnote{https://www.dreambank.net/}. Suppose Bob, a college student, would like to use \textit{Metamorpheus} to create a visual story based on his dream last night. He dreamt of walking down a beach, holding hands with a girl he used to like, so he creates a literal description as the opening scene. It says ``\textit{The dream began when I was walking along the beach with my old crush}''. He then wants to recreate the visual scene at this moment, which contains the beach, sunset, and his crush holding hands with himself. He therefore adds a metaphorical scene, and opens the metaphor editor, which prompted him to articulate what is affective, and describe it in one word.

\subsubsection{The First Scene: Old Crush Holding my Hands}
After some time of recall and reflection, he concludes that his old crush holding his hands was the most emotionally arousing stimulus, which made him feel excited at that moment. Therefore, he fills in the metaphor editor form, ``\textit{old crush holding my hands}'', and ``\textit{exciting}''. He then thinks about metaphors to draw, and wants to see suggestions first. He requests from ChatGPT words connected with, similar or opposite to the ``\textit{exciting}'' moment when \textit{old crush held his hands}. Suggestions include ``Electric Sparks'', ``Nostalgic Embrace'', ``Entangled Fingers'', etc. Bob accepts ``Electric Sparks'' as he thinks it is very accurate to describe the excitement. It also reminds him of a scene where her crush held his hands, surrounded by sparks. He therefore moves on to prompt editing, and chooses ``Fusion'' to update the prompt template in order to fuse her crush and himself hand in hand with ``Electric Sparks'' in one image.

After exploring for some time, Bob feels the generations were too random. Therefore he adds ``\textit{sunset on the beach}'' in the prompt. After regenerating, he becomes instantly intrigued by an image of one hand holding another lit up by sparks, as shown in \autoref{fig:example_1}. He really likes the atmosphere of the entire scene, which he thinks is perhaps warmed up by the colour of the sunset. He accepts the image, and feels that the default colour extracted from the image is pretty accurate to indicate the warmth of the scene. He also reads through the generated text depiction, which says that the ``\textit{sparks}'' ignite ``\textit{a thrilling sensation that surged through every inch of my being}''. Bob feels it was literary and to the point, but also reminds him of the next moment, where he sensed something else besides the thrill.

\subsubsection{The Final Scene: Hugging and Kissing}
Bob then moves on to creating another metaphorical scene, in which he suddenly realised that he had a girlfriend already, but he did not care, and hugged and kissed his old crush regardless. He contemplates a while over his emotions at that particular moment, and feels it was a mix of mostly thrill and a bit of worry that he might be caught cheating. He therefore opens the editor again and types ``\textit{hugging and kissing}'', and ``\textit{thrilling but a bit worrying}''. Bob browses through suggested metaphors, and then accepts ``Embracing Flames''. He thinks \textit{flames} can perfectly depict the thrilling moment, while implying a sense of destruction that can represent his worry. He then prompts the text-to-image model by both fusing the flames with kissing and hugging, and putting them in juxtaposition.

He finally accepts two images (see \autoref{fig:example_2}): one depicting them kissing and hugging so hard in front of fires to depict the thrill; the other one seemingly abstract with both warm and cold colours to depict a mix of thrill with worries. He then chooses reddish colours from both extracted palettes to represent the arousal level of his emotions. He also checks the generated text and feels that it perfectly explains his intended metaphorical meaning of \textit{flames}, saying that the ``\textit{passionate excitement}'' carries ``\textit{the risk of getting burned}''. To add a bit more context, he types at the beginning ``\textit{I realised I already had a girlfriend, but we kissed and hugged regardless.}''

Finally, Bob creates another literal scene as the ending, in which he types, ``\textit{The moment seemed so real that, when I woke up all of a sudden, I had thought it was reality, the thrill lingering on until I got up for coffee.}''. He then resizes the first scene a bit larger because he really likes the atmosphere, and relocates the second a bit above and towards the right end. Bob feels that in this way the story feels like building towards an inevitable end of destruction or disappear, as if the thrill can no longer exist after he wakes up.

\subsection{Implementation}
The user interface of \textit{Metamorpheus} is a web application implemented via React and JavaScript. The backend of \textit{Metamorpheus} serves a text-to-image model, Stable Diffusion \cite{rombach2022high}, that takes as input an incoming request with a text prompt, and returns a 512$\times$512 resolution image after 30 inference steps. The metaphor suggestions and metaphorical text depictions are generated by ChatGPT, i.e., GPT-3.5-turbo. The temperature was set to 1.0 for metaphor suggestions, and 0.7 for generating metaphorical depictions.

\input{bubbles}

%% file: scenario.tex
\begin{figure*}[htb]
\centering
\begin{minipage}{.64\textwidth}
    \centering
    \includegraphics[width=.95\linewidth]{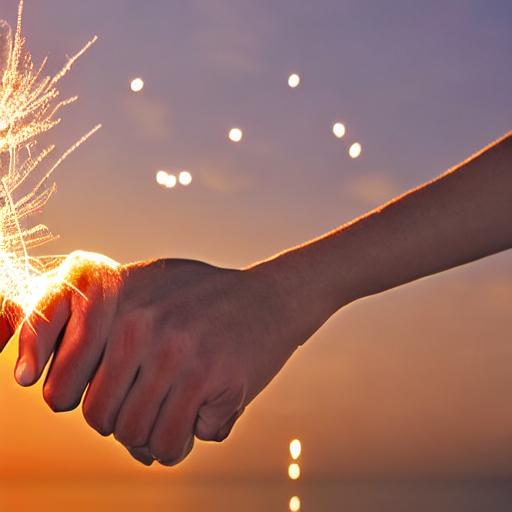}
    \subcaption{Scene \RNum{1}: Exciting Old crush holding my hands - Electric sparks}
    \label{fig:example_1}
\end{minipage}
\begin{minipage}{.32\textwidth}
    \centering\vspace{.05\linewidth}
    \includegraphics[width=.95\linewidth]{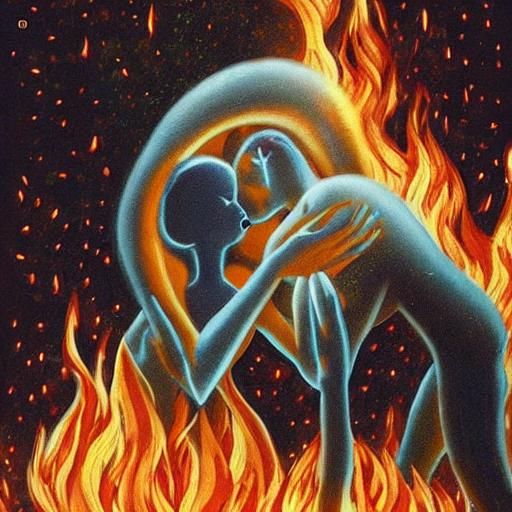}\vspace{.05\linewidth}
    \includegraphics[width=.95\linewidth]{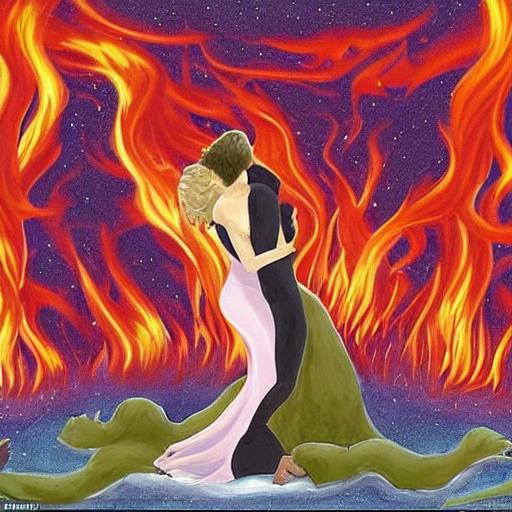}
    \subcaption{Scene \RNum{2}: Thrilling but a bit worrying Hugging and kissing - Embracing Flames}
    \label{fig:example_2}
\end{minipage}
\caption{Results of the example scenario}
\end{figure*}

%% file: bubbles.tex
\begin{figure}[!tb]
\centering
\includegraphics[width=\linewidth]{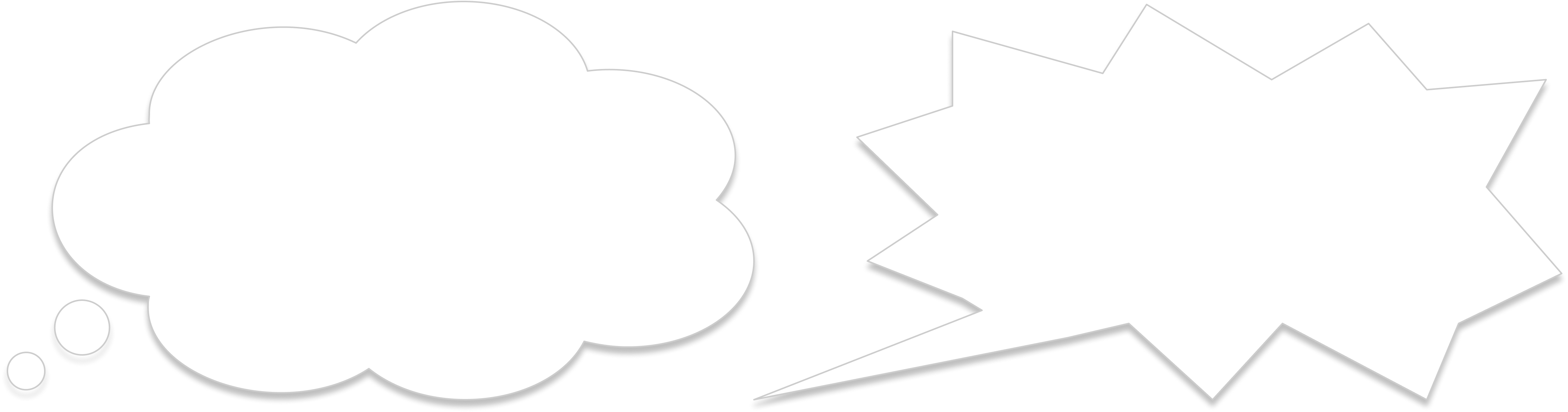}
\caption{Two different shapes for literal and metaphorical scenes.}
\label{fig:bubbles}
\end{figure}

%% file: 04-evaluation.tex
\section{Study Design}
The evaluation of \textit{Metamorpheus} was inspired by Höök's two-tiered design evaluation model for affective interfaces~\cite{hook2004user}, in which he advocated ``\textit{rich, narrative, constructive understanding}'' of interactions between users and affective systems. As we envisage the end goal of \textit{Metamorpheus} to be meaningful self-reflection, and its usage to be open-ended and scalable to other scenarios such as dream tracking or social sharing beyond an affective interface per se, we slightly modified the second level of the original two-tiered model, and hone in on the experience of meaningful reflection~\cite{mekler2019framework} throughout the interactions, as informed by Boehner et al.'s interactional model~\cite{boehner2005affect}.
Specifically, we attempted to answer the following questions:
\begin{itemize}
    \item \textbf{RQ1:} Are expressed emotions of \textit{Metamorpheus} readily understood, and do they correctly interpret users' emotions?
    \item \textbf{RQ2:} Does \textit{Metamorpheus} provoke meaningful reflection; and what are the meanings by users' accounts?
\end{itemize}
To evaluate separate features of \textit{Metamorpheus}, and obtain an in-situ understanding of users' experience of meaningful reflection, we conducted a lab study involving 12 participants. To answer \textbf{RQ1}, we conducted a survey after a story creation task to evaluate each key feature of the system.

To address \textbf{RQ2}, we took a phenomenological approach~\cite{moustakas1994phenomenological} and aimed to elicit and articulate the essence of experience during the interaction. In this sense, we did not set out to investigate our hypothesis of the presence of meaningful reflection in a simple yes or no question. Instead, we sought to obtain, as Höök put it, a ``\textit{constructive understanding}'' of how users experienced the interaction and co-creation results, of which the experience of meaning constitutes an essential component. To this end we prompted our participants to articulate their thoughts, feelings, perceptions, and understandings throughout the interaction, and conducted a semi-structured interview afterwards regarding the overall experience, including feelings, meaning made, sense of agency, etc. An interpretative phenomenological analysis were later performed on our qualitative data.

\subsection{Participants}
We recruited participants from our universities by advertising our user study on social media and through posters. In the end, 12 students (5 male, 7 female) aged between 18 and 30 agreed to participate in our study. For convenience we refer to them as P1-12.
Of the 12 participants, P10 disclosed that she had some experience of drawing, and P2, P11, \& P12 have a design background.
All participants were ethnically Chinese and L2 English speakers, and all studies were conducted in Mandarin. For each participant, we offered a 50 HKD coupon as an honorarium.

\subsection{Study Protocol}
Prior to the study, we obtained participants' consent to the use of their dream experiences by direct messaging, and asked them to recall 2-3 dreams from the past three months before they were invited to our lab. The lab study was later conducted in a private room where only one researcher and the participant were present. To build rapport with our participants, we first asked them about their attitudes towards dreams, whether they tracked their dreams, and their practices of dream tracking, if any. We then required participants to recount their experiences in each of their dreams, and recall their emotions while their narration unfolded.

After the dream narration, we demonstrated features of \textit{Metamorpheus} to participants by creating a visual story based on a randomly sampled dream narrative from Dream Bank. Participants were reminded that a creation from \textit{Metamorpheus} includes all visible elements on the user interface, such as text narrative, generated images, arrangement of the visualised storyline, applied colour filters, and so forth. After familiarising themselves with our system for 5 minutes, each participant was required to create a complete visual story based on one of their most affective dreams.

All participants were required to think aloud during the creation, and were frequently asked questions regarding their feelings, perceptions, and thought processes on the spot (e.g., why did you resize the event on the storyline? What did the results make you think? What do you feel about the colour filter?), especially when any non-verbal responses (e.g., chuckles, cringes, pondering, etc.) were observed. Because these experiences are often hard to describe, we asked participants multiple follow-up questions to reach an articulated interpretation (e.g., what made you think of this? what do you mean by feeling this way?). After the creation process concluded, we played the complete visual story scene by scene, and asked participants about their feelings, thoughts, and understandings.

In the end, we asked participants to complete a survey to rate three key features: generated visual metaphors, metaphorical text depictions, and colour filters. The survey questions are related to \textbf{RQ1}: whether expressed emotions of a feature were readily understood, and whether they correctly interpreted users' emotions. We also conducted a semi-structured interview with participants to reflect on their overall feelings, emotions, thought processes and mindfulness (e.g., how did the interaction or results impact your attitudes towards your emotions or dreams), sense of agency (e.g., how does the AI creation, or editing process affects your sense of agency in self-expression), meaning of their perceptions and understandings (during the think-aloud process), etc., during the interaction. The questions regarding the experience of meaning was informed by Mekler and Hornb{\ae}k's framework for the experience of meaning, and example questions across Mekler and Hornb{\ae}k's five dimensions can be found in \autoref{tab:interview}.

The whole process lasted between 1 and 1.5 hours, and was audio-taped, screen-recorded, and transcribed for further analysis.

\input{table-interview}

\subsection{Data Analysis}
The analysis of our qualitative study data was informed by the paradigm of interpretative phenomenological analysis~\cite{smith2007hermeneutics}. During the analysis, we aimed to cast out our hypothesis in \textbf{RQ2}, but answer it in the end with a rich narrative of how users made sense of the system, the intermediate and final results; how they constructed meaning out of the interaction, and what they thought of themselves during and after the interaction.

The qualitative data for analysis comprise screen-recording of the creation process, and transcriptions of the interview and think-aloud process. The first author, an HCI researcher that conducted all the studies, first performed an open coding~\cite{Corbin2014} of the transcriptions, while playing the screen recording simultaneously as a reference. Emerging codes were extracted and recorded with their timestamps in the screen recording, marked with participants' attidudes towards dreams, practices of dream recall, and initial dream narration at the very beginning. The initial codes that emerged were related to users' sense of agency, feelings of connection, motivations to share, thoughts on the workflow and creations, etc. We then performed a second round of coding to catalogue initial codes into themes regarding the essence of the experience of both the interactions and results. All data, codes, and themes were later translated into English for reporting.

\input{example}

%% file: table-interview.tex
\begin{table*}[htb]
    \centering
    \renewcommand{\arraystretch}{1.25}
    \resizebox{\textwidth}{!}{\begin{tabular}{ll}
    \toprule
    \textbf{Components of Meaning} & \textbf{Example Questions} \\
    \midrule
    \multirow{2}{*}{\textbf{Connectedness}} & \multirow{2}{*}{\makecell[l]{How did the use experience or results impact your view of your emotions, and anything related to your emotions? \\ Would you like to share these results with others? Why?}} \\
     & \\
    \multirow{2}{*}{\textbf{Purpose}} & \multirow{2}{*}{\makecell[l]{How might the system help discover new goals to strive for in your life? \\ How might the use experience impact or relate to your awareness of your emotions, dreams or anything related to them?}} \\
     & \\
    \textbf{Coherence} & Did the system help you discover anything related to yourself that you think was important or intriguing? \\
    \multirow{2}{*}{\textbf{Resonance}} & \multirow{2}{*}{\makecell[l]{What do you think of the final results of your creation? \\ Do they click with you emotionally? Any examples in particular?}} \\
     & \\
    \multirow{3}{*}{\textbf{Significance}} & \multirow{3}{*}{\makecell[l]{Do you think your interactions with the system are worthwhile? Why? \\ What are the benefits of using the system? \\ Under what circumstances will you consider using it again?}} \\
     & \\
     & \\
    \bottomrule
    \end{tabular}}
    \caption{Example interview questions across the five components of meaning in Mekler and Hornb{\ae}k's framework~\cite{mekler2019framework}}
    \label{tab:interview}
\end{table*}

%% file: example.tex
\begin{table*}[htb]
  \centering
  \renewcommand{\arraystretch}{0.15}
  \begin{tabular}{cm{2.75cm}m{6.25cm}l}
    \toprule
     \textbf{ID} & \textbf{Metaphor} & \textbf{Depiction} & \textbf{Image} \\
    \midrule
    P7 & Lively Banquet - Feast of Colors & The banquet was a feast of colors vibrant and alive It ignited a lively feeling within me like a canvas splashed with hues of joy and excitement Each dish a brushstroke creating a masterpiece of flavors that danced on my palate. & $\raisebox{-.5\height}{\includegraphics[width=0.18\linewidth]{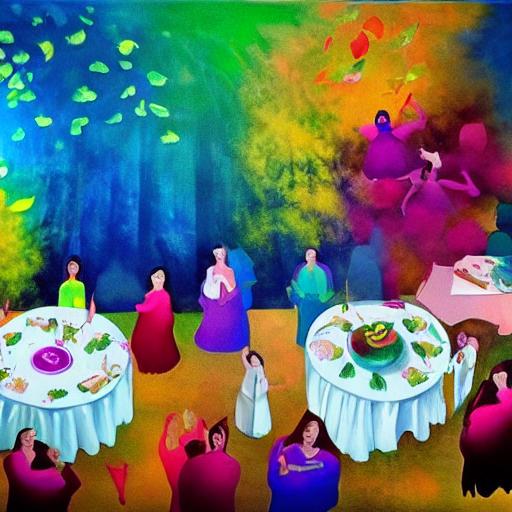}}$ \\
    & & & \\
    P8 & Beautiful Girl - Flower dance & Her presence blossoms in my heart like a graceful flower dance Each step she takes petals unfurl casting a spell on my soul The enchanting beauty she exudes is a melody that resonates within filling me with a profound indescribable joy. & $\raisebox{-.5\height}{\includegraphics[width=0.18\linewidth]{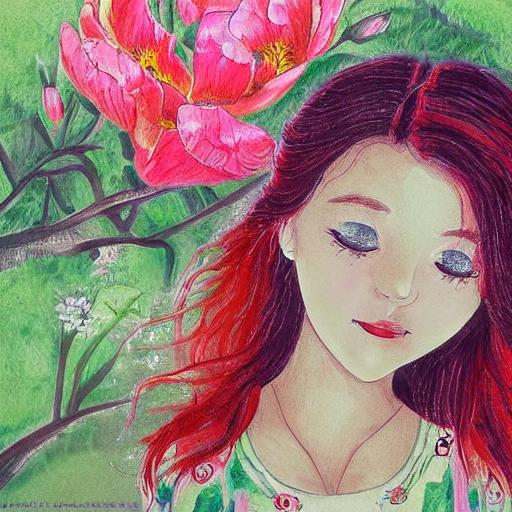}}$ \\
    & & & \\
    P8 & Moonlight Girl - Moonlit silhouette & I am enchanted by her presence like a moonlit silhouette that captivates the night Her essence shines upon my soul casting a gentle glow of love and wonder In her I find a celestial beauty that leaves me in a state of pure bliss. & $\raisebox{-.5\height}{\includegraphics[width=0.18\linewidth]{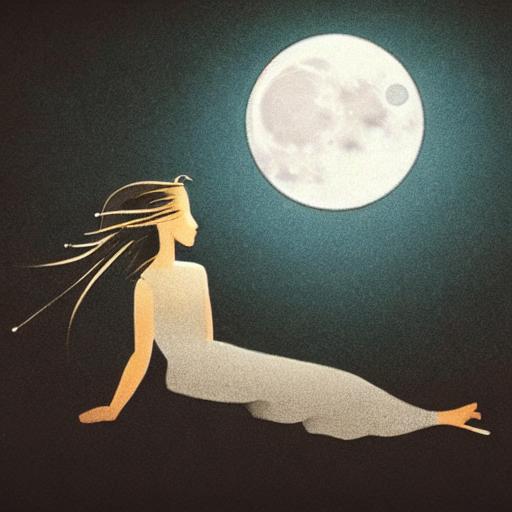}}$ \\
    & & & \\
    P11 & Disturbing Monster in the crowd - Bomb under the table & The "monster in the crowd" creates a chilling unease, much like a "bomb under the table." Both evoke a sense of imminent danger and the fear of the unknown, leaving a disturbing feeling that lingers in the depths of one's mind. & $\raisebox{-.5\height}{\includegraphics[width=0.18\linewidth]{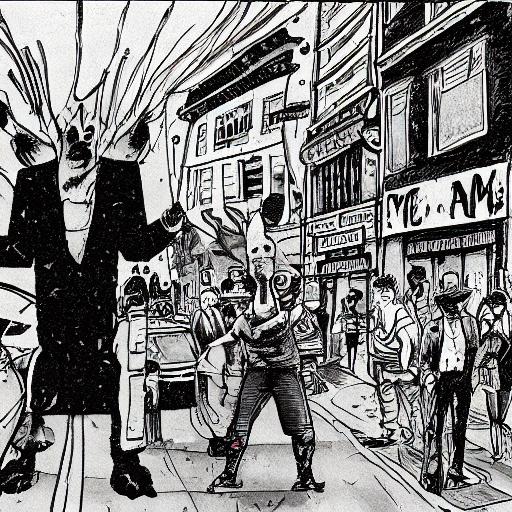}}$ \\
    & & & \\
    P12 & Joyful Friends - Bouquet of laughter & My friends are like a bouquet of laughter, filling my life with joy and happiness. Each friend adds a unique color and fragrance to my existence, creating a beautiful harmony that brings constant delight and warmth to my heart. & $\raisebox{-.5\height}{\includegraphics[width=0.18\linewidth]{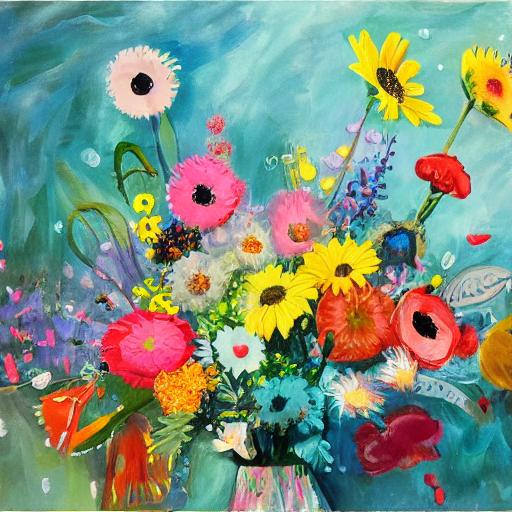}}$ \\
    & & & \\
    P12 & Boring Talking content - Text bubble drizzle & The "boring" feeling towards "talking content" can be likened to a drizzle of text bubbles. It lacks excitement or engagement, leaving one uninterested and unaffected.
    & $\raisebox{-.5\height}{\includegraphics[width=0.18\linewidth]{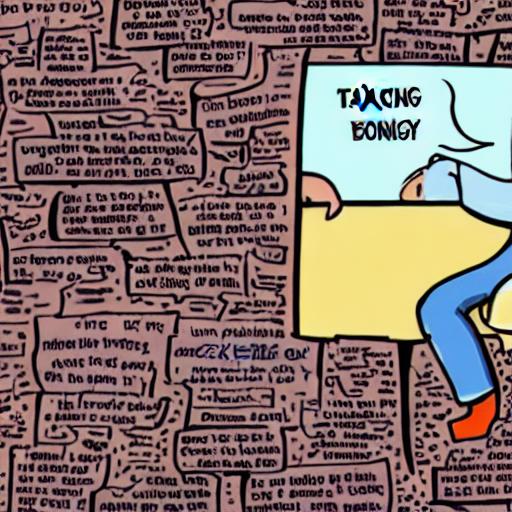}}$ \\
    \bottomrule
  \end{tabular}
  \caption{Example results of metaphorical scenes during our evaluation}
  \label{tab:example}
\end{table*}

%% file: 05-findings.tex
\section{Findings}
We report results of our study and tentatively answer our \textbf{RQ2} with a summary of the user experience of both the interaction and creation results.

\subsection{Feedback on Features: Accurate and Creative}
In general, participants regarded features of \textit{Metamorpheus} as accurate and easily understood, as shown in \autoref{fig:ablation}. There was only one case where the participant (P10) failed to obtain the desired generation during the study. In this case, P10 tried, but failed to recreate a very complex visual scene, where she ``\textit{held hands with her old friend and walked on white grounds, under the blue sky where white doves flew by, slowly passing arrays of white columns}''. She attributed the failure to either ``\textit{the complexity of the scene}'' or ``\textit{being poor at English and prompting}'', because she often turned to the researcher to help with the translation.

Besides image generation, all participants deemed metaphorical text depictions to accurately, though sometimes only partly (e.g, P5, P9, P11), reflected their emotions at the moment. The dominant colours extracted from the image were also deemed appropriate, and most participants found colour filters suitable for their emotions. There was only two cases where the participant (P6 \& P11) failed to find suitable colours in the extracted palette: In one case P6 felt the dominant brown colour was too dark and she preferred it with a grayish tint. She therefore input a hex colour code instead of using the colour picker. In the other case P11 needed a darker red to indicate a sense of lurking danger, when a monster was tailing him on a crowded street. He ended up choosing a grayish colour from the image to represent a disturbing mood because the image was predominantly gray.

We also found that participants came up with creative ways of arranging the visualisation. For instance, P2, P5, P7, and P11 positioned dangling images of each scene equally apart on the axis, and resized them according to the significance of events (P2, P7) or arousal levels of emotions (P5, P11), usually the higher the significance or arousal level, the larger the image. P3 instead resized each dangling image increasingly larger on the axis. She said this could build up the atmosphere in chronological order towards the final scene, where she finally found the culprit for her unfinished homework. P4 arranged dangling images of two scenes to become concentric circles that overlapped with each other. She explained that this could represent an overlap between past and present, which she believed was the central theme of her dream, where traumatic feelings from the past were re-enacted in her present life.

\input{ablation}

\subsection{Interactive Meaning Co-construction}
\sloppy{The interaction with \textit{Metamorpheus} used to be a meaning co-construction experience, in which the system facilitated self-expression, and participants further constructed meaning out of creative outputs in an open-ended manner based on their own affective experience.}

\subsubsection{Visual Metaphors as Meaning Co-construction}
Initially, most participants were found to recreate key visual details of their dream experiences via images of visual metaphors. For dreams that were still in vivid details, in particular, participants expected generated images to match their memories exactly. During this period, the text-to-image model often failed participants in that it was too random for correctly recreating an abstract visual scene during a dream. For instance, P11 described that he was looking for key elements in his memories such as ``\textit{a very crowded street}'', ``\textit{a monster hidden in the crowd}'', ``\textit{being tailed by the monster that felt like an imminent danger}''. The generated images kept failing him for lacking either crowds, monsters, or a sense of danger.

Nevertheless, after some time of exploring, participants usually ended up constructing their personalised meanings out of random outputs, and looking for patterns similar to their memories, such as colours, overall atmosphere, certain objects, or certain visual patterns, instead of perfect recreations. For example, in P12's dreams, he was chatting with old friends over dinner, and gradually getting bored by their conversation. He used \textit{bouquet of laughter} to represent his friends, and \textit{text bubble drizzle} to represent their conversation. Instead of recreating visual details, he accepted two images for these two scenes that seemed rather abstract, even without any traces of his friends in human form (see \autoref{tab:example}). He explained that the first image accurately recreated ``\textit{the joyful atmosphere}'', and the second image felt like ``\textit{someone cramped in these text bubbles, and trying to escape}''.

P8 was recreating visual scenes involving his old crush. He later provided a detailed explanation of what he expected from the text-to-image model during the interview,
\begin{quote}
    ``\textit{I cared about whether the overall feelings (of the image) matched (my feelings) towards her in the dream. It's a common sense the model can't recreate a portrait of her... What I primarily expected, or perhaps cared about, was first, whether the surroundings or atmosphere fit; second, a rough recreation of her figure; third, whether the overall image resonates with me emotionally.}''
\end{quote}
He further pointed to two images he accepted in his final visual story, as shown in~\autoref{tab:example}, and explained that the first image of his old crush had ``\textit{melancholic eyes}'', which reflected his ``\textit{sense of regret}''. The second image did not contain a clear picture of her face, but rather visualises her figure as sitting next to the moon, lit by moonlight. He said it made him feel she was pretty and perfect like the moon, but ``\textit{distant and out of reach}''.

\subsubsection{Text Depictions for Meaning Co-construction}
After the image generation, most participants (P8-9, P11-12) reported that the generated metaphorical text depictions helped articulate their emotions in a literary way. Perhaps because our participants were all non-native English speakers, some of them (P3, P6, P11) said the depictions were very vivid and elegant, and beyond their own of English writing skills.

For example, P3 drew a metaphor between an \textit{empty backpack} and her \textit{unfinished homework} to depict a horrified feeling, for which ChatGPT generated,
\begin{quote}
    Unfinished homework feels like carrying an empty backpack devoid of purpose and weight. It hangs heavy on my mind a constant reminder of unfulfilled responsibilities. The emptiness echoes a nagging reminder of the task yet to be completed
\end{quote}
She felt descriptions such as ``\textit{hangs heavy on my mind}'' and ``\textit{a constant reminder of unfulfilled responsibilities}'' were quite vivid and to the point, which she might find difficult to articulate.
While depicting a scene of a unexpected escape from a building, P6 accepted a suggested metaphor \textit{vanishing footsteps}, and ChatGPT later generated, ``\textit{My unexpected feeling towards escaping is like vanishing footsteps leaving no trace behind.}'' She later commented during the interview that the metaphor was very graphic, and perfectly depicted a quite abstract visual scene, which was beyond her imagination at that time.


\subsubsection{Alternating Agency to Describe the Ineffable}
During the interaction, the agency was alternating between users and the system in the meaning co-construction process. The system, especially the generative AI, took part in the self-expression, while users constructed new meaning and iterated the expression again out of the system outputs. When asked if they felt this co-creation took away from them the agency of expression, all participants almost unanimously disagreed. Only P9 and P10 said it might depend.

P4 noted that without the system, especially the process of metaphor editing and the visualisation of the affect-driven storyline, she would not even know how to visualise or narrate the dream, because dreams and emotions in dreams were ``\textit{too abstract}''. P12 also mentioned that the workflow of the system served as an appropriate way of instructing him on how to develop the story. During the interview, P6 reflected that,
\begin{quote}
    ``\textit{I used to track my dreams in a very informal and aimless way. It might be too random. This form of expression kind of drives me to articulate by creating a visual story that matches my feelings in the dream... I felt that the workflow of metaphor editing acted as an effective device, by which I could articulate my feelings. Without it I might not know how to express these abstract feelings.}''
\end{quote}

As the only participant with some experience of drawing, P10 said that she might want to draw her dream scenes if they were too abstract or complex for text-to-image models to recreate. However, she generally preferred using the system first because it would be much more convenient, and its workflow could help recall details in the dream. P9 shared the same opinion that he would turn to drawing by himself if his dreams happened to be too hard to recreate. He preferred using the system because a free-form creation such as drawing ``\textit{might seem distracting and complicated for expressing emotions or dreams}'', and ``\textit{does not necessarily mean it can reflect my emotions more accurately}''

\subsection{The Interaction Relives the Dream in Detail}
All our participants almost unanimously reported that the interaction with \textit{Metamorpheus} was an experience of reliving all details in a dream, especially related to their own emotions.

\subsubsection{Mindfulness through Reflection}
The interaction with \textit{Metamorpheus} brought alive emotional experiences in a dream. Participants described it as an ``\textit{re-enactment}'' (P2), a ``\textit{reminder}'' (P1), an ``\textit{amplification}'' (P3), a ``\textit{reinforcement}'', and a ``\textit{detailed and vivid recollection}'' (P4, P6, P11) of past emotions. P11 said he did not even notice his dreams were actually full of emotions until he used the system. Some others (e.g., P5-7, P9) shared similar views, that they previously had a habit of dream tracking, but did not do it with a focus on emotions. They generally agreed that this interaction raised their awareness of their emotions, as they believed dreams reflected part of their waking thoughts. P6 recalled during the interview that,
\begin{quote}
    ``\textit{sometimes I experienced something in a dream, but I'm now out of the dream and perhaps unable to articulate them. The system can offer creative and even random suggestions in various forms to remind me of them in the dream... It also asked me to recall and reflect on my emotions in the dream... which reinforced my memories of them.}''
\end{quote}

We have also found cases where \textit{Metamorpheus} helped discover hidden emotional details in a dream that participants had almost forgotten. For instance, P7 dreamt about a wedding banquet, where she was getting married and blessed by friends. During the interview, she said in retrospect,
\begin{quote}
    ``\textit{I feel I now have a better understanding of my emotions at that moment. I thought I was mainly shy during the dream, and trying to tell my friends not to make a big fuss. But while I was recalling the moment and creating the visual story, I felt a great sense of happiness... I think the sense of happiness used to be hidden in my memories, but now is activated... Especially the colours of the last image, they are lively and cheerful, which reinforces that feeling of happiness.}'' 
\end{quote}

Similarly, P9 discovered new interpretations of his emotions in a monochrome dream. The ChatGPT generations interpreted one monochrome scene as \textit{a mix of contrasting emotions, where the absence of colour makes everything appear more nuanced and subtle, creating a captivating and timeless atmosphere}. P9 said he did not feel the contrasting emotions no matter how he tried to recall, but he took it as an interpretation of his subconsciousness. While this depicted a feeling he did not experience, he still found it informative, as it raised awareness of potential emotions deep within his subconscious self.

\subsubsection{A Reflective but Innocuous Manner}
We found that, though our participants felt it challenging to articulate, the experience of reliving and reflecting on the emotional experiences was generally benign.
The reported effects even resembled those found in expressive therapies~\cite{malchiodi2013expressive}, though participants such as P4 refused to use the word ``\textit{therapeutic}''. Instead, the interaction was described as an experience that ``\textit{felt like roasting from a distance}'' (P3), one with ``\textit{a sense of detachment}'' (P4, P9) and ``\textit{abstract expression instead of concrete re-creation}'' (P8), and as if ``\textit{in a third-person perspective}'' (P3), and ``\textit{in a peaceful and comfortable manner}'' (P4), which raised their awareness of their own emotions, including negative ones, in an innocuous manner.

For example, P4 was creating a visual story based on a nightmare related to her traumatic experiences. She said the interaction helped her relive her memories, but she described the experience as an ``\textit{escape}''. While creating the visual story, she felt as if she ``\textit{escaped from the experience}'' and relived these memories ``\textit{with a sense of detachment}'', which was quite ``\textit{peaceful and comfortable}''. She reasoned that, ``\textit{I felt I was more focused on the creation and putting emotion into words, which perhaps attenuated my emotions}''. P3 also came across negative feelings in the middle of her dream. She described the interaction experience as ``\textit{roasting}'' herself in the dream by re-creating a visual story, and ``\textit{at a distance}'' from that particular feeling, which did not make her feel uncomfortable.

P9 said that he took generations of the system with a grain of salt, especially the metaphorical text depictions of his so-called ``\textit{subconscious emotions}''. By drawing an analogy with dream dictionaries, he said that, ``\textit{Whether it (the depiction) was accurate I'm not sure. It's like a dream dictionary that interprets my dreams. It might say I will become rich if I dream of something, but that's it. Whether I will truly become rich I don't know.}'' Therefore, P9 added that he would generally treat the interpretation merely as a reference.

\subsection{Connectedness, and Motivations to Share}
We found that the interaction with \textit{Metamorpheus} and its outcome created a sense of connectedness, and motivations to share dream experiences. To our surprise, P3 and P4 almost immediately asked the researcher if they could keep these results for sharing, before we were about to ask them whether they wanted to. P3 said she would like to use one image as her profile picture, because it was ``\textit{so funny}'', and ``\textit{perfectly reflected a delighted mood}''. She also noted that the overall experience was very ``\textit{funny}'' and ``\textit{engaging}'', and she particularly enjoyed conversing with the researcher during the creation process, which she said ``\textit{provided more ideas and inspiration, and amplified my happiness}''. She added that she hoped to use this system with her friends together, if possible, in the future.

P4 instead was creating a visual story of a nightmare. She mentioned that she wanted to share these results with her friends that had similar traumatic experiences. She explained that,
\begin{quote}
    ``\textit{Whenever I woke up from a dream like this, I would immediately share with my friends that had the same experience. I would text them and say, Oh, I dreamt of it again. I felt that sharing with them these results is a more, emmm, milder way than a pure text narration. Cause sometimes I just don't want to disturb them with my emotions, but I felt this way of sharing is quite, emmm, metaphorical.}''
\end{quote}
She also added that she felt using this system with her close friends would be more engaging and meaningful.

Besides, many participants (P1, P3-5, P7-9, P11) mentioned that these results of creation, including both visual metaphors and metaphorical text depictions, can serve as a better way of sharing, as compared to traditional text or speech narratives that tend to be lengthy or tedious for communication. P4 and P5 said that with images of the story arranged chronologically in a visualisation, it would be much more ``\textit{appealing}'' and ``\textit{easier to follow}'' than a long piece of text if they ever wanted to share with others. P8 said that images and the entire interface are more ``\textit{intuitive}'' so that others could learn about what he dreamt and what he might feel during the dream at first glance.

We also noticed some participants came up with diverse and creative ways of sharing these results. P11 mentioned that he felt dreams might be associated with creativity, and the creation of metaphors was very creative. As a designer, he might share these results with his friends or colleagues to draw inspirations, and perhaps encourage them to also use this system to discover creativity in dreams together. P2 dreamt of being a character in the game \textit{The Legend of Zelda}. She said she wanted to share the visual stories with other players and let them guess the game. She might also use creation results to communicate with developers to improve a game because they contain visual scenes and storylines.

\subsection{Summary}
We present the summary of our findings, in an answer to our \textbf{RQ2}. We found that \textit{Metamorpheus} was an effective way of meaning co-construction between users and computers, which was accurate and creative. The interaction did provoke self-reflection full of meaning, in that
\begin{enumerate}
\item it engaged users in a meaning making process, where creative outputs of the system prompted users to construct personalised meaning.
\item it promoted mindfulness of emotions by vividly reliving affective experiences in a dream, but in an innocuous manner, with a therapy-like effect.
\item it created a sense of connectedness, including the connectedness between users and observers, and the motivations for dream sharing in a more creative and efficient way.
\end{enumerate}

%% file: ablation.tex
\begin{figure*}[ht]
    \centering
    \includegraphics[width=.85\linewidth]{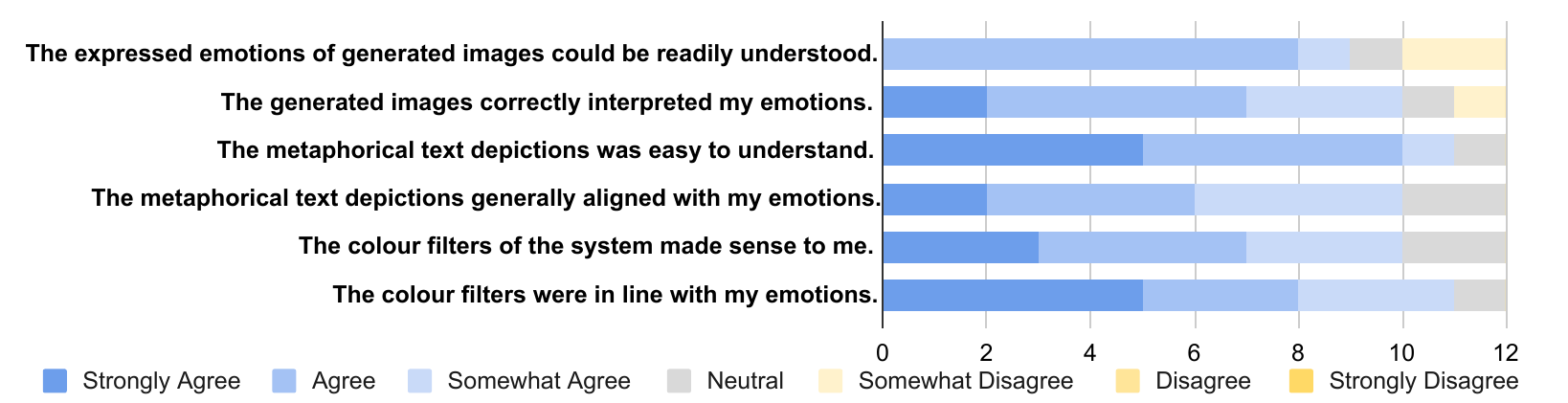}
    \caption{Feedback on \textit{Metamorpheus} features.}
    \label{fig:ablation}
\end{figure*}

%% file: 06-discussion.tex
\section{Discussion}
\textit{Metamorpheus} features the paradigm of co-creation of visual stories as a way of meaning co-construction. It complements previous studies of personal informatics and affective computing, and opens up the avenue for computer-supported creative meaning co-construction to support the practice of self-tracking, digital journaling, or even therapies.
\subsection{Revisiting Agency and Ownership in the Co-construction of Meaning}
Statistical models of emotions, such as emotion recognition algorithms, are both pervasive in our real life, and dominant in the discourse of affective computing. Previous studies criticised them for impeding users' sense of agency~\cite{andalibi2020human}, and objectifying and dehumanizing users~\cite{chancellor2019human}.
Focusing on the lived experiences of users, our study results suggest that the co-creative workflow of \textit{Metamorpheus} affords a sense of agency, as participants were able to craft their individualised expression by co-creating images and text depictions, re-arranging storyline visualisations, and applying colour filters to recolour the interface. Participants also took ownership of these results, as they have constructed personalised meaning, even out of those automatically generated.
These findings in general align with previous studies of paper bullet journals~\cite{tholander2020crafting,ayobi2018flexible}, and collectively prove that our design goal has been achieved at large.
The design of \textit{Metamorpheus} is a step forward from previous affective systems such as Rajcic and McCormack's \textit{Mirror Ritual}~\cite{rajcic2020mirror}: not only machines can co-construct meanings with users by creative generations, but users themselves are now able to participate in the creation as a self-expression process, and informs the creation interactively with their lived-experiences. The additional design features beyond brute AI generation or mere image co-creation also addressed the concern of an \textit{overpowering} AI~\cite{du2024deepthink} in an envisioned human-AI art therapy session.

Nevertheless, it is worth noting that, compared to other non-digital methods of creative expression in the wild (e.g., bullet journaling~\cite{tholander2020crafting,ayobi2018flexible}), our system does guide users to recall their emotions, use metaphors for creation, and arrange the storyline according to emotions in a linear order, which is indeed reflected in our two somewhat contradicting design goals. In this workflow, our participants generally did not report any loss of \textit{control}, \textit{autonomy}, \textit{agency}, or any related sentiments. Instead, their accounts suggest that guidance is needed to articulate the ineffable emotional experiences they encounter in dreams, which are often complex and abstract. This finding needs to be taken with skepticism since few of our users have an expertise in art, painting, or literature, etc.
Only one of our participant (P10) with some experience of drawing said the creation was much more convenient for memory reactivation with the guidance provided by \textit{Metamorpheus}, but it is far from conclusive for all expert users.
Yet it should be safe to say that future design of digital creative narration needs to consider some forms of guidance, scaffolding, or suggestions to facilitate both creation and recollection of ineffable experiences such as dreams~\cite{hoefer2022personal}.

\subsection{Design for Affective Experience: Generative AI \& Phenomenology}
\textit{Metamorpheus} contributes to the field of experience-centred design~\cite{wright2008aesthetics,mccarthy2004technology} an affective interface to support creative narration of dream experiences. It reveals the potential of generative AI models for meaning-making in the recollection of affective, and perhaps lived experiences in general, to promote mindfulness and provoke reflection. Our study even suggests it might have the potential of creating empathy~\cite{bennett2019promise} between the researcher and the user (e.g., P3). In this process, AI continuously provides creative and often random outputs for users to construct personalised meaning. This finding challenges the traditional role of AI as a surveillance-style detector~\cite{andalibi2020human} that made users feel invasive and scary. We look forward to applications of generative AI in computing affective experiences by interactive meaning co-construction beyond creative narration. For instance, how can generative AI help connect with others emotionally in a broader socio-technical context; how can it help relate our emotions to past experiences as a means of memory reactivation or self-reflection~\cite{wright2008aesthetics}?

Furthermore, the phenomenological methodology in our study also provides an alternative to measuring both experience of meaning~\cite{mekler2019framework}, and use experience of interacting with affective interfaces~\cite{hook2004user,boehner2005affect}. Phenomenology was used mainly because we reasoned that the evaluation of our system, especially \textbf{RQ2}, should be better studied by \textit{bracketing} the in-situ experience of the interaction, the intermediate, and the final results, suspending our prior knowledge and hypothesis. Similar to other experiences such as dementia~\cite{morrissey2017value} that are hard to encapsulate, dreams were also found to contain ineffable emotions. In a response to this, we conducted a think-aloud lab study so that we can attend to users' non-verbal response and require them to articulate their experience on-site. Compared to the widely used micro-phenomenology~\cite{prpa2020articulating} in experience-centred design, our approach is more interpretative in that 1) our questions during the interview and the think-aloud study were informed by design goals, hypothesis, previous theories and literature; 2) the data were interpreted by referring to participants' attitudes towards dreams, practices of dream recall, and initial versions of narration.

\subsection{Future Work: Co-creative Narration as Meaning Co-construction}
The design of \textit{Metamorpheus} probes into the potential of introducing co-creative narration as a process of meaning co-construction. We expect that diverse paradigms across various modalities can be developed in the future to facilitate creative narration of lived experiences. For example, similar to ~\cite{chung2022talebrush}, we can introduce pen and touch interactions and enable users to create their personal stories based on the valence or arousal level of emotions. 
The device of metaphor creation should also be considered, as it proved to be creative and engaging, and could put the ineffable into words. It has also been used as an efficient way of communication in today's clinical practices (e.g., describing pain~\cite{stilwell2021painful}). In the context of digital journaling, therefore, we can design to engage users in metaphorical recollection or narration of personal experiences to facilitate meaning making.
To keep the interaction creative and open-ended to support agency, similar to our second design goal, we can further consider allowing users to iterate co-creative results, or alter or re-arrange the interface using co-creative results.

The design of co-creative narration systems can also inspire future work in personal informatics. We expect computer-supported creative narration be deployed to facilitate tracking of life experiences including but not limited to dreams. The device of metaphor creation in our system might also enrich existing personal informatics tools.
As we accidentally found out the metaphor creation process built empathy and rapport between users and researchers, it is reasonable to consider using the metaphor creation process to facilitate sharing of personal data without invading privacy. For instance, personal informatics systems can consider engaging multiple users in metaphor creation to exchange their personal information such as health, diet, emotions, etc., without explicitly disclosing raw data.
Besides, as we found that creative narration often motivates and facilitates sharing, similar features or workflows can also be deployed on social media platforms to facilitate disclosure or sharing of personalised life events~\cite{saha2021life}.

Furthermore, the computer-supported creative narration also has the potential of bringing about therapeutic benefits. Our findings suggest that, with proper guidance, generative AI models are able to help articulate the often ineffable emotional experiences in a dream, while retaining the agency of the user. We expect this can inform future design to support users in addressing mental health conditions that are hard to describe. Social media platforms, for example, can offer AI co-creative narration systems to support users with traumatic experiences~\cite{chen2022trauma,scott2023trauma}.

\subsection{Limitations}
There are several key limitations of this paper that arise from our study method. First, the phenomenological approach heavily relies on the participants' capabilities to articulate their experiences. This is particularly challenging in dream experiences as they are often hard to describe. In this sense, the data collected might be too superficial to reach the real essence of participants' experiences.
Second, all of our participants were ethnically Chinese and L2 English speakers. Therefore when prompting the ChatGPT or Stable Diffusion in English, they might not be able to fully articulate their emotions, and sometimes fail to draw an appropriate metaphor. Their perception of suggested metaphors, and generated metaphorical text depictions might also differ from a proficient or native English speaker.
Furthermore, most of our participants were non-expert in drawing or art. Therefore our findings might not directly generalise to those with expertise. It also remains uncertain the design or artistic background of some of our participants might impact the results.

%% file: 07-conclusion.tex
\section{Conclusion}
In this paper, we present the design of \textit{Metamorpheus}, an affective interface that engages users in the creative narration of emotional experiences in a dream. \textit{Metamorpheus} features the co-creation of visual stories as a means of meaning co-construction, and the creation results includes images of visual metaphors, text depictions, storyline visualisation, and colour filters. Our phenomenological evaluation suggests that interacting with \textit{Metamorpheus} was meaningful in that it promotes mindfulness and creates connectedness in an innocuous manner.

%% file: main.bbl

\begin{thebibliography}{92}


\ifx \showCODEN    \undefined \def \showCODEN     #1{\unskip}     \fi
\ifx \showDOI      \undefined \def \showDOI       #1{#1}\fi
\ifx \showISBNx    \undefined \def \showISBNx     #1{\unskip}     \fi
\ifx \showISBNxiii \undefined \def \showISBNxiii  #1{\unskip}     \fi
\ifx \showISSN     \undefined \def \showISSN      #1{\unskip}     \fi
\ifx \showLCCN     \undefined \def \showLCCN      #1{\unskip}     \fi
\ifx \shownote     \undefined \def \shownote      #1{#1}          \fi
\ifx \showarticletitle \undefined \def \showarticletitle #1{#1}   \fi
\ifx \showURL      \undefined \def \showURL       {\relax}        \fi
\providecommand\bibfield[2]{#2}
\providecommand\bibinfo[2]{#2}
\providecommand\natexlab[1]{#1}
\providecommand\showeprint[2][]{arXiv:#2}

\bibitem[Adams(1999)]%
        {Adams1999}
\bibfield{author}{\bibinfo{person}{Kathleen Adams}.} \bibinfo{year}{1999}\natexlab{}.
\newblock \bibinfo{booktitle}{\emph{A Brief History of Journal Writing}}.
\newblock
\urldef\tempurl%
\url{https://web.archive.org/web/20130205033640/http://journaltherapy.com/journaltherapy/journal-cafe/journal-writing-history}
\showURL{%
Retrieved Aug 18, 2023 from \tempurl}


\bibitem[Andalibi and Buss(2020)]%
        {andalibi2020human}
\bibfield{author}{\bibinfo{person}{Nazanin Andalibi} {and} \bibinfo{person}{Justin Buss}.} \bibinfo{year}{2020}\natexlab{}.
\newblock \showarticletitle{The human in emotion recognition on social media: Attitudes, outcomes, risks}. In \bibinfo{booktitle}{\emph{Proceedings of the 2020 CHI Conference on Human Factors in Computing Systems}}. \bibinfo{pages}{1--16}.
\newblock


\bibitem[Aoki et~al\mbox{.}(2022)]%
        {aoki2022emoballoon}
\bibfield{author}{\bibinfo{person}{Toshiki Aoki}, \bibinfo{person}{Rintaro Chujo}, \bibinfo{person}{Katsufumi Matsui}, \bibinfo{person}{Saemi Choi}, {and} \bibinfo{person}{Ari Hautasaari}.} \bibinfo{year}{2022}\natexlab{}.
\newblock \showarticletitle{Emoballoon-conveying emotional arousal in text chats with speech balloons}. In \bibinfo{booktitle}{\emph{Proceedings of the 2022 CHI Conference on Human Factors in Computing Systems}}. \bibinfo{pages}{1--16}.
\newblock


\bibitem[Ayobi et~al\mbox{.}(2020)]%
        {ayobi2020trackly}
\bibfield{author}{\bibinfo{person}{Amid Ayobi}, \bibinfo{person}{Paul Marshall}, {and} \bibinfo{person}{Anna~L Cox}.} \bibinfo{year}{2020}\natexlab{}.
\newblock \showarticletitle{Trackly: A customisable and pictorial self-tracking app to support agency in multiple sclerosis self-care}. In \bibinfo{booktitle}{\emph{Proceedings of the 2020 CHI Conference on Human Factors in Computing Systems}}. \bibinfo{pages}{1--15}.
\newblock


\bibitem[Ayobi et~al\mbox{.}(2018)]%
        {ayobi2018flexible}
\bibfield{author}{\bibinfo{person}{Amid Ayobi}, \bibinfo{person}{Tobias Sonne}, \bibinfo{person}{Paul Marshall}, {and} \bibinfo{person}{Anna~L Cox}.} \bibinfo{year}{2018}\natexlab{}.
\newblock \showarticletitle{Flexible and mindful self-tracking: Design implications from paper bullet journals}. In \bibinfo{booktitle}{\emph{Proceedings of the 2018 CHI conference on human factors in computing systems}}. \bibinfo{pages}{1--14}.
\newblock


\bibitem[Barrett(2006)]%
        {barrett2006solving}
\bibfield{author}{\bibinfo{person}{Lisa~Feldman Barrett}.} \bibinfo{year}{2006}\natexlab{}.
\newblock \showarticletitle{Solving the emotion paradox: Categorization and the experience of emotion}.
\newblock \bibinfo{journal}{\emph{Personality and social psychology review}} \bibinfo{volume}{10}, \bibinfo{number}{1} (\bibinfo{year}{2006}), \bibinfo{pages}{20--46}.
\newblock


\bibitem[Barrett(2017a)]%
        {barrett2017emotions}
\bibfield{author}{\bibinfo{person}{Lisa~Feldman Barrett}.} \bibinfo{year}{2017}\natexlab{a}.
\newblock \bibinfo{booktitle}{\emph{How emotions are made: The secret life of the brain}}.
\newblock \bibinfo{publisher}{Pan Macmillan}.
\newblock


\bibitem[Barrett(2017b)]%
        {barrett2017theory}
\bibfield{author}{\bibinfo{person}{Lisa~Feldman Barrett}.} \bibinfo{year}{2017}\natexlab{b}.
\newblock \showarticletitle{The theory of constructed emotion: an active inference account of interoception and categorization}.
\newblock \bibinfo{journal}{\emph{Social cognitive and affective neuroscience}} \bibinfo{volume}{12}, \bibinfo{number}{1} (\bibinfo{year}{2017}), \bibinfo{pages}{1--23}.
\newblock


\bibitem[Barrett et~al\mbox{.}(2007)]%
        {barrett2007mice}
\bibfield{author}{\bibinfo{person}{Lisa~Feldman Barrett}, \bibinfo{person}{Kristen~A Lindquist}, \bibinfo{person}{Eliza Bliss-Moreau}, \bibinfo{person}{Seth Duncan}, \bibinfo{person}{Maria Gendron}, \bibinfo{person}{Jennifer Mize}, {and} \bibinfo{person}{Lauren Brennan}.} \bibinfo{year}{2007}\natexlab{}.
\newblock \showarticletitle{Of mice and men: Natural kinds of emotions in the mammalian brain? A response to Panksepp and Izard}.
\newblock \bibinfo{journal}{\emph{Perspectives on psychological science}} \bibinfo{volume}{2}, \bibinfo{number}{3} (\bibinfo{year}{2007}), \bibinfo{pages}{297--312}.
\newblock


\bibitem[Barrett and Russell(2014)]%
        {barrett2014psychological}
\bibfield{author}{\bibinfo{person}{Lisa~Feldman Barrett} {and} \bibinfo{person}{James~A Russell}.} \bibinfo{year}{2014}\natexlab{}.
\newblock \bibinfo{booktitle}{\emph{The psychological construction of emotion}}.
\newblock \bibinfo{publisher}{Guilford Publications}.
\newblock


\bibitem[Barrett and Satpute(2013)]%
        {barrett2013large}
\bibfield{author}{\bibinfo{person}{Lisa~Feldman Barrett} {and} \bibinfo{person}{Ajay~Bhaskar Satpute}.} \bibinfo{year}{2013}\natexlab{}.
\newblock \showarticletitle{Large-scale brain networks in affective and social neuroscience: towards an integrative functional architecture of the brain}.
\newblock \bibinfo{journal}{\emph{Current opinion in neurobiology}} \bibinfo{volume}{23}, \bibinfo{number}{3} (\bibinfo{year}{2013}), \bibinfo{pages}{361--372}.
\newblock


\bibitem[Bell and Hall(2011)]%
        {bell2011personality}
\bibfield{author}{\bibinfo{person}{Alan~Paul Bell} {and} \bibinfo{person}{Calvin~Springer Hall}.} \bibinfo{year}{2011}\natexlab{}.
\newblock \bibinfo{booktitle}{\emph{The personality of a child molester: An analysis of dreams}}.
\newblock \bibinfo{publisher}{Transaction Publishers}.
\newblock


\bibitem[Bennett and Rosner(2019)]%
        {bennett2019promise}
\bibfield{author}{\bibinfo{person}{Cynthia~L Bennett} {and} \bibinfo{person}{Daniela~K Rosner}.} \bibinfo{year}{2019}\natexlab{}.
\newblock \showarticletitle{The promise of empathy: Design, disability, and knowing the" other"}. In \bibinfo{booktitle}{\emph{Proceedings of the 2019 CHI conference on human factors in computing systems}}. \bibinfo{pages}{1--13}.
\newblock


\bibitem[Boals(2012)]%
        {boals2012use}
\bibfield{author}{\bibinfo{person}{Adriel Boals}.} \bibinfo{year}{2012}\natexlab{}.
\newblock \showarticletitle{The use of meaning making in expressive writing: When meaning is beneficial}.
\newblock \bibinfo{journal}{\emph{Journal of Social and Clinical Psychology}} \bibinfo{volume}{31}, \bibinfo{number}{4} (\bibinfo{year}{2012}), \bibinfo{pages}{393--409}.
\newblock


\bibitem[Boehner et~al\mbox{.}(2005)]%
        {boehner2005affect}
\bibfield{author}{\bibinfo{person}{Kirsten Boehner}, \bibinfo{person}{Rog{\'e}rio DePaula}, \bibinfo{person}{Paul Dourish}, {and} \bibinfo{person}{Phoebe Sengers}.} \bibinfo{year}{2005}\natexlab{}.
\newblock \showarticletitle{Affect: from information to interaction}. In \bibinfo{booktitle}{\emph{Proceedings of the 4th decennial conference on Critical computing: between sense and sensibility}}. \bibinfo{pages}{59--68}.
\newblock


\bibitem[Cartwright et~al\mbox{.}(2006)]%
        {cartwright2006relation}
\bibfield{author}{\bibinfo{person}{Rosalind Cartwright}, \bibinfo{person}{Mehmet~Y Agargun}, \bibinfo{person}{Jennifer Kirkby}, {and} \bibinfo{person}{Julie~Kabat Friedman}.} \bibinfo{year}{2006}\natexlab{}.
\newblock \showarticletitle{Relation of dreams to waking concerns}.
\newblock \bibinfo{journal}{\emph{Psychiatry research}} \bibinfo{volume}{141}, \bibinfo{number}{3} (\bibinfo{year}{2006}), \bibinfo{pages}{261--270}.
\newblock


\bibitem[Chancellor et~al\mbox{.}(2019)]%
        {chancellor2019human}
\bibfield{author}{\bibinfo{person}{Stevie Chancellor}, \bibinfo{person}{Eric~PS Baumer}, {and} \bibinfo{person}{Munmun De~Choudhury}.} \bibinfo{year}{2019}\natexlab{}.
\newblock \showarticletitle{Who is the" human" in human-centered machine learning: The case of predicting mental health from social media}.
\newblock \bibinfo{journal}{\emph{Proceedings of the ACM on Human-Computer Interaction}} \bibinfo{volume}{3}, \bibinfo{number}{CSCW} (\bibinfo{year}{2019}), \bibinfo{pages}{1--32}.
\newblock


\bibitem[Chang et~al\mbox{.}(2015)]%
        {chang2015palette}
\bibfield{author}{\bibinfo{person}{Huiwen Chang}, \bibinfo{person}{Ohad Fried}, \bibinfo{person}{Yiming Liu}, \bibinfo{person}{Stephen DiVerdi}, {and} \bibinfo{person}{Adam Finkelstein}.} \bibinfo{year}{2015}\natexlab{}.
\newblock \showarticletitle{Palette-based photo recoloring.}
\newblock \bibinfo{journal}{\emph{ACM Trans. Graph.}} \bibinfo{volume}{34}, \bibinfo{number}{4} (\bibinfo{year}{2015}), \bibinfo{pages}{139--1}.
\newblock


\bibitem[Chen et~al\mbox{.}(2022)]%
        {chen2022trauma}
\bibfield{author}{\bibinfo{person}{Janet~X Chen}, \bibinfo{person}{Allison McDonald}, \bibinfo{person}{Yixin Zou}, \bibinfo{person}{Emily Tseng}, \bibinfo{person}{Kevin~A Roundy}, \bibinfo{person}{Acar Tamersoy}, \bibinfo{person}{Florian Schaub}, \bibinfo{person}{Thomas Ristenpart}, {and} \bibinfo{person}{Nicola Dell}.} \bibinfo{year}{2022}\natexlab{}.
\newblock \showarticletitle{Trauma-informed computing: Towards safer technology experiences for all}. In \bibinfo{booktitle}{\emph{Proceedings of the 2022 CHI conference on human factors in computing systems}}. \bibinfo{pages}{1--20}.
\newblock


\bibitem[Chung et~al\mbox{.}(2022)]%
        {chung2022talebrush}
\bibfield{author}{\bibinfo{person}{John Joon~Young Chung}, \bibinfo{person}{Wooseok Kim}, \bibinfo{person}{Kang~Min Yoo}, \bibinfo{person}{Hwaran Lee}, \bibinfo{person}{Eytan Adar}, {and} \bibinfo{person}{Minsuk Chang}.} \bibinfo{year}{2022}\natexlab{}.
\newblock \showarticletitle{TaleBrush: Sketching stories with generative pretrained language models}. In \bibinfo{booktitle}{\emph{Proceedings of the 2022 CHI Conference on Human Factors in Computing Systems}}. \bibinfo{pages}{1--19}.
\newblock


\bibitem[Corbin and Strauss(2014)]%
        {Corbin2014}
\bibfield{author}{\bibinfo{person}{Juliet Corbin} {and} \bibinfo{person}{Anselm Strauss}.} \bibinfo{year}{2014}\natexlab{}.
\newblock \bibinfo{booktitle}{\emph{Basics of qualitative research: Techniques and procedures for developing grounded theory}}.
\newblock \bibinfo{publisher}{Sage publications}.
\newblock


\bibitem[de~Castle and Mind(1994)]%
        {de1994ballantine}
\bibfield{author}{\bibinfo{person}{Van de Castle} {and} \bibinfo{person}{RL~Our~Dreaming Mind}.} \bibinfo{year}{1994}\natexlab{}.
\newblock \showarticletitle{Ballantine Books}.
\newblock \bibinfo{journal}{\emph{New York}} (\bibinfo{year}{1994}).
\newblock


\bibitem[Domhoff(1996)]%
        {domhoff1996finding}
\bibfield{author}{\bibinfo{person}{G~William Domhoff}.} \bibinfo{year}{1996}\natexlab{}.
\newblock \showarticletitle{Finding meaning in dreams: A quantitative approach}.
\newblock  (\bibinfo{year}{1996}).
\newblock


\bibitem[Domhoff(2017)]%
        {domhoff2017invasion}
\bibfield{author}{\bibinfo{person}{G~William Domhoff}.} \bibinfo{year}{2017}\natexlab{}.
\newblock \showarticletitle{The invasion of the concept snatchers: The origins, distortions, and future of the continuity hypothesis.}
\newblock \bibinfo{journal}{\emph{Dreaming}} \bibinfo{volume}{27}, \bibinfo{number}{1} (\bibinfo{year}{2017}), \bibinfo{pages}{14}.
\newblock


\bibitem[Du et~al\mbox{.}(2024)]%
        {du2024deepthink}
\bibfield{author}{\bibinfo{person}{Xuejun Du}, \bibinfo{person}{Pengcheng An}, \bibinfo{person}{Justin Leung}, \bibinfo{person}{April Li}, \bibinfo{person}{Linda~E Chapman}, {and} \bibinfo{person}{Jian Zhao}.} \bibinfo{year}{2024}\natexlab{}.
\newblock \showarticletitle{DeepThInk: Designing and probing human-AI co-creation in digital art therapy}.
\newblock \bibinfo{journal}{\emph{International Journal of Human-Computer Studies}}  \bibinfo{volume}{181} (\bibinfo{year}{2024}), \bibinfo{pages}{103139}.
\newblock


\bibitem[Ekman(1992)]%
        {ekman1992argument}
\bibfield{author}{\bibinfo{person}{Paul Ekman}.} \bibinfo{year}{1992}\natexlab{}.
\newblock \showarticletitle{An argument for basic emotions}.
\newblock \bibinfo{journal}{\emph{Cognition \& emotion}} \bibinfo{volume}{6}, \bibinfo{number}{3-4} (\bibinfo{year}{1992}), \bibinfo{pages}{169--200}.
\newblock


\bibitem[Ekman et~al\mbox{.}(1999)]%
        {ekman1999basic}
\bibfield{author}{\bibinfo{person}{Paul Ekman} {et~al\mbox{.}}} \bibinfo{year}{1999}\natexlab{}.
\newblock \showarticletitle{Basic emotions}.
\newblock \bibinfo{journal}{\emph{Handbook of cognition and emotion}} \bibinfo{volume}{98}, \bibinfo{number}{45-60} (\bibinfo{year}{1999}), \bibinfo{pages}{16}.
\newblock


\bibitem[Elsden et~al\mbox{.}(2016)]%
        {elsden2016s}
\bibfield{author}{\bibinfo{person}{Chris Elsden}, \bibinfo{person}{Abigail~C Durrant}, {and} \bibinfo{person}{David~S Kirk}.} \bibinfo{year}{2016}\natexlab{}.
\newblock \showarticletitle{It's just my history isn't it? Understanding smart journaling practices}. In \bibinfo{booktitle}{\emph{Proceedings of the 2016 CHI Conference on Human Factors in Computing Systems}}. \bibinfo{pages}{2819--2831}.
\newblock


\bibitem[Evans(2017)]%
        {evans2017emoji}
\bibfield{author}{\bibinfo{person}{Vyvyan Evans}.} \bibinfo{year}{2017}\natexlab{}.
\newblock \bibinfo{booktitle}{\emph{The emoji code: How smiley faces, love hearts and thumbs up are changing the way we communicate}}.
\newblock \bibinfo{publisher}{Michael O'Mara Books}.
\newblock


\bibitem[Fleshman and Fryrear(1981)]%
        {fleshman1981arts}
\bibfield{author}{\bibinfo{person}{Bob Fleshman} {and} \bibinfo{person}{Jerry~L Fryrear}.} \bibinfo{year}{1981}\natexlab{}.
\newblock \showarticletitle{The arts in therapy}.
\newblock \bibinfo{journal}{\emph{(No Title)}} (\bibinfo{year}{1981}).
\newblock


\bibitem[Forceville(2011)]%
        {forceville2011pictorial}
\bibfield{author}{\bibinfo{person}{Charles Forceville}.} \bibinfo{year}{2011}\natexlab{}.
\newblock \showarticletitle{Pictorial runes in Tintin and the Picaros}.
\newblock \bibinfo{journal}{\emph{Journal of Pragmatics}} \bibinfo{volume}{43}, \bibinfo{number}{3} (\bibinfo{year}{2011}), \bibinfo{pages}{875--890}.
\newblock


\bibitem[Gendron and Barrett(2018)]%
        {gendron2018emotion}
\bibfield{author}{\bibinfo{person}{Maria Gendron} {and} \bibinfo{person}{Lisa~Feldman Barrett}.} \bibinfo{year}{2018}\natexlab{}.
\newblock \showarticletitle{Emotion perception as conceptual synchrony}.
\newblock \bibinfo{journal}{\emph{Emotion Review}} \bibinfo{volume}{10}, \bibinfo{number}{2} (\bibinfo{year}{2018}), \bibinfo{pages}{101--110}.
\newblock


\bibitem[Gero and Chilton(2019)]%
        {gero2019metaphoria}
\bibfield{author}{\bibinfo{person}{Katy~Ilonka Gero} {and} \bibinfo{person}{Lydia~B Chilton}.} \bibinfo{year}{2019}\natexlab{}.
\newblock \showarticletitle{Metaphoria: An algorithmic companion for metaphor creation}. In \bibinfo{booktitle}{\emph{Proceedings of the 2019 CHI conference on human factors in computing systems}}. \bibinfo{pages}{1--12}.
\newblock


\bibitem[Gieselmann et~al\mbox{.}(2019)]%
        {gieselmann2019aetiology}
\bibfield{author}{\bibinfo{person}{Annika Gieselmann}, \bibinfo{person}{Malik Ait~Aoudia}, \bibinfo{person}{Michelle Carr}, \bibinfo{person}{Anne Germain}, \bibinfo{person}{Robert Gorzka}, \bibinfo{person}{Brigitte Holzinger}, \bibinfo{person}{Birgit Kleim}, \bibinfo{person}{Barry Krakow}, \bibinfo{person}{Anna~E Kunze}, \bibinfo{person}{Jaap Lancee}, {et~al\mbox{.}}} \bibinfo{year}{2019}\natexlab{}.
\newblock \showarticletitle{Aetiology and treatment of nightmare disorder: State of the art and future perspectives}.
\newblock \bibinfo{journal}{\emph{Journal of sleep research}} \bibinfo{volume}{28}, \bibinfo{number}{4} (\bibinfo{year}{2019}), \bibinfo{pages}{e12820}.
\newblock


\bibitem[Givrad(2016)]%
        {givrad2016dream}
\bibfield{author}{\bibinfo{person}{Soudabeh Givrad}.} \bibinfo{year}{2016}\natexlab{}.
\newblock \showarticletitle{Dream theory and science: A review}.
\newblock \bibinfo{journal}{\emph{Psychoanalytic Inquiry}} \bibinfo{volume}{36}, \bibinfo{number}{3} (\bibinfo{year}{2016}), \bibinfo{pages}{199--213}.
\newblock


\bibitem[Gladding(1992)]%
        {gladding1992counseling}
\bibfield{author}{\bibinfo{person}{Samuel~T Gladding}.} \bibinfo{year}{1992}\natexlab{}.
\newblock \bibinfo{booktitle}{\emph{Counseling as an art: The creative arts in counseling.}}
\newblock \bibinfo{publisher}{ERIC}.
\newblock


\bibitem[Hall and Van~de Castle(1966)]%
        {hall1966content}
\bibfield{author}{\bibinfo{person}{Calvin Hall} {and} \bibinfo{person}{Robert Van~de Castle}.} \bibinfo{year}{1966}\natexlab{}.
\newblock \showarticletitle{The content analysis of dreams}.
\newblock  (\bibinfo{year}{1966}).
\newblock


\bibitem[Halliday(1987)]%
        {halliday1987direct}
\bibfield{author}{\bibinfo{person}{Gordon Halliday}.} \bibinfo{year}{1987}\natexlab{}.
\newblock \showarticletitle{Direct psychological therapies for nightmares: A review}.
\newblock \bibinfo{journal}{\emph{Clinical psychology review}} \bibinfo{volume}{7}, \bibinfo{number}{5} (\bibinfo{year}{1987}), \bibinfo{pages}{501--523}.
\newblock


\bibitem[Halperin and Lukin(2023)]%
        {halperin2023envisioning}
\bibfield{author}{\bibinfo{person}{Brett~A Halperin} {and} \bibinfo{person}{Stephanie~M Lukin}.} \bibinfo{year}{2023}\natexlab{}.
\newblock \showarticletitle{Envisioning Narrative Intelligence: A Creative Visual Storytelling Anthology}. In \bibinfo{booktitle}{\emph{Proceedings of the 2023 CHI Conference on Human Factors in Computing Systems}}. \bibinfo{pages}{1--21}.
\newblock


\bibitem[Harrison et~al\mbox{.}(2007)]%
        {harrison2007three}
\bibfield{author}{\bibinfo{person}{Steve Harrison}, \bibinfo{person}{Deborah Tatar}, {and} \bibinfo{person}{Phoebe Sengers}.} \bibinfo{year}{2007}\natexlab{}.
\newblock \showarticletitle{The three paradigms of HCI}. In \bibinfo{booktitle}{\emph{Alt. Chi. Session at the SIGCHI Conference on human factors in computing systems San Jose, California, USA}}. \bibinfo{pages}{1--18}.
\newblock


\bibitem[Hill(1996)]%
        {hill1996working}
\bibfield{author}{\bibinfo{person}{Clara~E Hill}.} \bibinfo{year}{1996}\natexlab{}.
\newblock \bibinfo{booktitle}{\emph{Working with dreams in psychotherapy}}.
\newblock \bibinfo{publisher}{Guilford Press}.
\newblock


\bibitem[Hill et~al\mbox{.}(1997)]%
        {hill1997dream}
\bibfield{author}{\bibinfo{person}{Clara~E Hill}, \bibinfo{person}{Roberta~A Diemer}, {and} \bibinfo{person}{Kristin~J Heaton}.} \bibinfo{year}{1997}\natexlab{}.
\newblock \showarticletitle{Dream interpretation sessions: Who volunteers, who benefits, and what volunteer clients view as most and least helpful.}
\newblock \bibinfo{journal}{\emph{Journal of Counseling Psychology}} \bibinfo{volume}{44}, \bibinfo{number}{1} (\bibinfo{year}{1997}), \bibinfo{pages}{53}.
\newblock


\bibitem[Hoefer et~al\mbox{.}(2022)]%
        {hoefer2022personal}
\bibfield{author}{\bibinfo{person}{Michael Jeffrey~Daniel Hoefer}, \bibinfo{person}{Bryce~E Schumacher}, {and} \bibinfo{person}{Stephen Voida}.} \bibinfo{year}{2022}\natexlab{}.
\newblock \showarticletitle{Personal Dream Informatics: A Self-Information Systems Model of Dream Engagement}. In \bibinfo{booktitle}{\emph{Proceedings of the 2022 CHI Conference on Human Factors in Computing Systems}}. \bibinfo{pages}{1--16}.
\newblock


\bibitem[Hoel(2021)]%
        {hoel2021overfitted}
\bibfield{author}{\bibinfo{person}{Erik Hoel}.} \bibinfo{year}{2021}\natexlab{}.
\newblock \showarticletitle{The overfitted brain: Dreams evolved to assist generalization}.
\newblock \bibinfo{journal}{\emph{Patterns}} \bibinfo{volume}{2}, \bibinfo{number}{5} (\bibinfo{year}{2021}).
\newblock


\bibitem[H{\"o}{\"o}k(2004)]%
        {hook2004user}
\bibfield{author}{\bibinfo{person}{Kristina H{\"o}{\"o}k}.} \bibinfo{year}{2004}\natexlab{}.
\newblock \showarticletitle{User-centred design and evaluation of affective interfaces: A two-tiered model}.
\newblock \bibinfo{journal}{\emph{From brows to trust: evaluating embodied conversational agents}} (\bibinfo{year}{2004}), \bibinfo{pages}{127--160}.
\newblock


\bibitem[Kang et~al\mbox{.}(2021)]%
        {kang2021metamap}
\bibfield{author}{\bibinfo{person}{Youwen Kang}, \bibinfo{person}{Zhida Sun}, \bibinfo{person}{Sitong Wang}, \bibinfo{person}{Zeyu Huang}, \bibinfo{person}{Ziming Wu}, {and} \bibinfo{person}{Xiaojuan Ma}.} \bibinfo{year}{2021}\natexlab{}.
\newblock \showarticletitle{MetaMap: Supporting visual metaphor ideation through multi-dimensional example-based exploration}. In \bibinfo{booktitle}{\emph{Proceedings of the 2021 CHI Conference on Human Factors in Computing Systems}}. \bibinfo{pages}{1--15}.
\newblock


\bibitem[Keller et~al\mbox{.}(1995)]%
        {keller1995use}
\bibfield{author}{\bibinfo{person}{John~W Keller}, \bibinfo{person}{Gina Brown}, \bibinfo{person}{Katja Maier}, \bibinfo{person}{Korinne Steinfurth}, \bibinfo{person}{Shelly Hall}, {and} \bibinfo{person}{Chris Piotrowski}.} \bibinfo{year}{1995}\natexlab{}.
\newblock \showarticletitle{Use of dreams in therapy: A survey of clinicians in private practice}.
\newblock \bibinfo{journal}{\emph{Psychological Reports}} \bibinfo{volume}{76}, \bibinfo{number}{3\_suppl} (\bibinfo{year}{1995}), \bibinfo{pages}{1288--1290}.
\newblock


\bibitem[Krakow et~al\mbox{.}(1995)]%
        {krakow1995imagery}
\bibfield{author}{\bibinfo{person}{Barry Krakow}, \bibinfo{person}{Robert Kellner}, \bibinfo{person}{Dorothy Pathak}, {and} \bibinfo{person}{Lori Lambert}.} \bibinfo{year}{1995}\natexlab{}.
\newblock \showarticletitle{Imagery rehearsal treatment for chronic nightmares}.
\newblock \bibinfo{journal}{\emph{Behaviour Research and Therapy}} \bibinfo{volume}{33}, \bibinfo{number}{7} (\bibinfo{year}{1995}), \bibinfo{pages}{837--843}.
\newblock


\bibitem[Krakow and Zadra(2006)]%
        {krakow2006clinical}
\bibfield{author}{\bibinfo{person}{Barry Krakow} {and} \bibinfo{person}{Antonio Zadra}.} \bibinfo{year}{2006}\natexlab{}.
\newblock \showarticletitle{Clinical management of chronic nightmares: imagery rehearsal therapy}.
\newblock \bibinfo{journal}{\emph{Behavioral sleep medicine}} \bibinfo{volume}{4}, \bibinfo{number}{1} (\bibinfo{year}{2006}), \bibinfo{pages}{45--70}.
\newblock


\bibitem[Kuss{\'e} et~al\mbox{.}(2010)]%
        {kusse2010neuroimaging}
\bibfield{author}{\bibinfo{person}{Caroline Kuss{\'e}}, \bibinfo{person}{Vincenzo Muto}, \bibinfo{person}{Laura Mascetti}, \bibinfo{person}{Luca Matarazzo}, \bibinfo{person}{Ariane Foret}, \bibinfo{person}{Anahita Shaffii-Le~Bourdiec}, {and} \bibinfo{person}{Pierre Maquet}.} \bibinfo{year}{2010}\natexlab{}.
\newblock \showarticletitle{Neuroimaging of dreaming: state of the art and limitations}.
\newblock \bibinfo{journal}{\emph{International Review of Neurobiology}}  \bibinfo{volume}{92} (\bibinfo{year}{2010}), \bibinfo{pages}{87--99}.
\newblock


\bibitem[Lieberman et~al\mbox{.}(2007)]%
        {lieberman2007affect}
\bibfield{author}{\bibinfo{person}{Matthew~D Lieberman}, \bibinfo{person}{Naomi~I Eisenberger}, \bibinfo{person}{Molly~J Crockett}, \bibinfo{person}{Sabrina~M Tom}, \bibinfo{person}{Jennifer~H Pfeifer}, {and} \bibinfo{person}{BM Way}.} \bibinfo{year}{2007}\natexlab{}.
\newblock \showarticletitle{Affect labeling disrupts amygdala activity in response to affective stimuli}.
\newblock \bibinfo{journal}{\emph{Psychological Science}} \bibinfo{volume}{18}, \bibinfo{number}{5} (\bibinfo{year}{2007}), \bibinfo{pages}{421--428}.
\newblock


\bibitem[Lindquist et~al\mbox{.}(2012)]%
        {lindquist2012brain}
\bibfield{author}{\bibinfo{person}{Kristen~A Lindquist}, \bibinfo{person}{Tor~D Wager}, \bibinfo{person}{Hedy Kober}, \bibinfo{person}{Eliza Bliss-Moreau}, {and} \bibinfo{person}{Lisa~Feldman Barrett}.} \bibinfo{year}{2012}\natexlab{}.
\newblock \showarticletitle{The brain basis of emotion: a meta-analytic review}.
\newblock \bibinfo{journal}{\emph{Behavioral and brain sciences}} \bibinfo{volume}{35}, \bibinfo{number}{3} (\bibinfo{year}{2012}), \bibinfo{pages}{121--143}.
\newblock


\bibitem[Ma et~al\mbox{.}(2022)]%
        {ma2022glancee}
\bibfield{author}{\bibinfo{person}{Shuai Ma}, \bibinfo{person}{Taichang Zhou}, \bibinfo{person}{Fei Nie}, {and} \bibinfo{person}{Xiaojuan Ma}.} \bibinfo{year}{2022}\natexlab{}.
\newblock \showarticletitle{Glancee: An adaptable system for instructors to grasp student learning status in synchronous online classes}. In \bibinfo{booktitle}{\emph{Proceedings of the 2022 CHI Conference on Human Factors in Computing Systems}}. \bibinfo{pages}{1--25}.
\newblock


\bibitem[Malchiodi(2013)]%
        {malchiodi2013expressive}
\bibfield{author}{\bibinfo{person}{Cathy~A Malchiodi}.} \bibinfo{year}{2013}\natexlab{}.
\newblock \bibinfo{booktitle}{\emph{Expressive therapies}}.
\newblock \bibinfo{publisher}{Guilford Publications}.
\newblock


\bibitem[Maratos et~al\mbox{.}(2008)]%
        {maratos2008music}
\bibfield{author}{\bibinfo{person}{Anna Maratos}, \bibinfo{person}{Christian Gold}, \bibinfo{person}{Xu Wang}, {and} \bibinfo{person}{Mike Crawford}.} \bibinfo{year}{2008}\natexlab{}.
\newblock \showarticletitle{Music therapy for depression}.
\newblock \bibinfo{journal}{\emph{Cochrane database of systematic reviews}} \bibinfo{number}{1} (\bibinfo{year}{2008}).
\newblock


\bibitem[McCarthy and Wright(2004)]%
        {mccarthy2004technology}
\bibfield{author}{\bibinfo{person}{John McCarthy} {and} \bibinfo{person}{Peter Wright}.} \bibinfo{year}{2004}\natexlab{}.
\newblock \showarticletitle{Technology as experience}.
\newblock \bibinfo{journal}{\emph{interactions}} \bibinfo{volume}{11}, \bibinfo{number}{5} (\bibinfo{year}{2004}), \bibinfo{pages}{42--43}.
\newblock


\bibitem[McNiff(1981)]%
        {mcniff1981arts}
\bibfield{author}{\bibinfo{person}{Shaun McNiff}.} \bibinfo{year}{1981}\natexlab{}.
\newblock \bibinfo{booktitle}{\emph{The arts and psychotherapy}}.
\newblock \bibinfo{publisher}{Charles C. Thomas}.
\newblock


\bibitem[McNiff(1992)]%
        {mcniff1992art}
\bibfield{author}{\bibinfo{person}{Shaun McNiff}.} \bibinfo{year}{1992}\natexlab{}.
\newblock \bibinfo{booktitle}{\emph{Art as medicine: Creating a therapy of the imagination}}.
\newblock \bibinfo{publisher}{Shambhala Publications}.
\newblock


\bibitem[Mekler and Hornb{\ae}k(2019)]%
        {mekler2019framework}
\bibfield{author}{\bibinfo{person}{Elisa~D Mekler} {and} \bibinfo{person}{Kasper Hornb{\ae}k}.} \bibinfo{year}{2019}\natexlab{}.
\newblock \showarticletitle{A framework for the experience of meaning in human-computer interaction}. In \bibinfo{booktitle}{\emph{Proceedings of the 2019 CHI conference on human factors in computing systems}}. \bibinfo{pages}{1--15}.
\newblock


\bibitem[Morrissey et~al\mbox{.}(2017)]%
        {morrissey2017value}
\bibfield{author}{\bibinfo{person}{Kellie Morrissey}, \bibinfo{person}{John McCarthy}, {and} \bibinfo{person}{Nadia Pantidi}.} \bibinfo{year}{2017}\natexlab{}.
\newblock \showarticletitle{The value of experience-centred design approaches in dementia research contexts}. In \bibinfo{booktitle}{\emph{Proceedings of the 2017 CHI Conference on Human Factors in Computing Systems}}. \bibinfo{pages}{1326--1338}.
\newblock


\bibitem[Moustakas(1994)]%
        {moustakas1994phenomenological}
\bibfield{author}{\bibinfo{person}{Clark Moustakas}.} \bibinfo{year}{1994}\natexlab{}.
\newblock \bibinfo{booktitle}{\emph{Phenomenological research methods}}.
\newblock \bibinfo{publisher}{Sage publications}.
\newblock


\bibitem[Murali et~al\mbox{.}(2021)]%
        {murali2021affectivespotlight}
\bibfield{author}{\bibinfo{person}{Prasanth Murali}, \bibinfo{person}{Javier Hernandez}, \bibinfo{person}{Daniel McDuff}, \bibinfo{person}{Kael Rowan}, \bibinfo{person}{Jina Suh}, {and} \bibinfo{person}{Mary Czerwinski}.} \bibinfo{year}{2021}\natexlab{}.
\newblock \showarticletitle{Affectivespotlight: Facilitating the communication of affective responses from audience members during online presentations}. In \bibinfo{booktitle}{\emph{Proceedings of the 2021 CHI Conference on Human Factors in Computing Systems}}. \bibinfo{pages}{1--13}.
\newblock


\bibitem[Ostriker(2018)]%
        {ostriker2018poetry}
\bibfield{author}{\bibinfo{person}{Alicia Ostriker}.} \bibinfo{year}{2018}\natexlab{}.
\newblock \showarticletitle{Poetry and Healing: Some Moments of Wholeness}.
\newblock \bibinfo{journal}{\emph{The American Poetry Review}} \bibinfo{volume}{47}, \bibinfo{number}{2} (\bibinfo{year}{2018}), \bibinfo{pages}{9--12}.
\newblock


\bibitem[Palagini and Rosenlicht(2011)]%
        {palagini2011sleep}
\bibfield{author}{\bibinfo{person}{Laura Palagini} {and} \bibinfo{person}{Nicholas Rosenlicht}.} \bibinfo{year}{2011}\natexlab{}.
\newblock \showarticletitle{Sleep, dreaming, and mental health: a review of historical and neurobiological perspectives}.
\newblock \bibinfo{journal}{\emph{Sleep Medicine Reviews}} \bibinfo{volume}{15}, \bibinfo{number}{3} (\bibinfo{year}{2011}), \bibinfo{pages}{179--186}.
\newblock


\bibitem[Pearson and Wilson(2008)]%
        {pearson2008using}
\bibfield{author}{\bibinfo{person}{Mark Pearson} {and} \bibinfo{person}{Helen Wilson}.} \bibinfo{year}{2008}\natexlab{}.
\newblock \showarticletitle{Using expressive counselling tools to enhance emotional literacy, emotional wellbeing and resilience: Improving therapeutic outcomes with Expressive Therapies}.
\newblock \bibinfo{journal}{\emph{Counselling, Psychotherapy and Health}} \bibinfo{volume}{4}, \bibinfo{number}{1} (\bibinfo{year}{2008}), \bibinfo{pages}{1--19}.
\newblock


\bibitem[Pennebaker(1997)]%
        {pennebaker1997writing}
\bibfield{author}{\bibinfo{person}{James~W Pennebaker}.} \bibinfo{year}{1997}\natexlab{}.
\newblock \showarticletitle{Writing about emotional experiences as a therapeutic process}.
\newblock \bibinfo{journal}{\emph{Psychological science}} \bibinfo{volume}{8}, \bibinfo{number}{3} (\bibinfo{year}{1997}), \bibinfo{pages}{162--166}.
\newblock


\bibitem[Petrelli et~al\mbox{.}(2009)]%
        {petrelli2009making}
\bibfield{author}{\bibinfo{person}{Daniela Petrelli}, \bibinfo{person}{Elise Van~den Hoven}, {and} \bibinfo{person}{Steve Whittaker}.} \bibinfo{year}{2009}\natexlab{}.
\newblock \showarticletitle{Making history: intentional capture of future memories}. In \bibinfo{booktitle}{\emph{Proceedings of the SIGCHI conference on Human Factors in computing systems}}. \bibinfo{pages}{1723--1732}.
\newblock


\bibitem[Phillips and McQuarrie(2004)]%
        {phillips2004beyond}
\bibfield{author}{\bibinfo{person}{Barbara~J Phillips} {and} \bibinfo{person}{Edward~F McQuarrie}.} \bibinfo{year}{2004}\natexlab{}.
\newblock \showarticletitle{Beyond visual metaphor: A new typology of visual rhetoric in advertising}.
\newblock \bibinfo{journal}{\emph{Marketing theory}} \bibinfo{volume}{4}, \bibinfo{number}{1-2} (\bibinfo{year}{2004}), \bibinfo{pages}{113--136}.
\newblock


\bibitem[Picard(2000)]%
        {picard2000affective}
\bibfield{author}{\bibinfo{person}{Rosalind~W Picard}.} \bibinfo{year}{2000}\natexlab{}.
\newblock \bibinfo{booktitle}{\emph{Affective computing}}.
\newblock \bibinfo{publisher}{MIT press}.
\newblock


\bibitem[Pifalo(2007)]%
        {pifalo2007jogging}
\bibfield{author}{\bibinfo{person}{Terry Pifalo}.} \bibinfo{year}{2007}\natexlab{}.
\newblock \showarticletitle{Jogging the cogs: Trauma-focused art therapy and cognitive behavioral therapy with sexually abused children}.
\newblock \bibinfo{journal}{\emph{Art Therapy}} \bibinfo{volume}{24}, \bibinfo{number}{4} (\bibinfo{year}{2007}), \bibinfo{pages}{170--175}.
\newblock


\bibitem[Prpa et~al\mbox{.}(2020)]%
        {prpa2020articulating}
\bibfield{author}{\bibinfo{person}{Mirjana Prpa}, \bibinfo{person}{Sarah Fdili-Alaoui}, \bibinfo{person}{Thecla Schiphorst}, {and} \bibinfo{person}{Philippe Pasquier}.} \bibinfo{year}{2020}\natexlab{}.
\newblock \showarticletitle{Articulating experience: Reflections from experts applying micro-phenomenology to design research in HCI}. In \bibinfo{booktitle}{\emph{Proceedings of the 2020 CHI Conference on Human Factors in Computing Systems}}. \bibinfo{pages}{1--14}.
\newblock


\bibitem[Rajcic and McCormack(2020a)]%
        {rajcic2020mirror}
\bibfield{author}{\bibinfo{person}{Nina Rajcic} {and} \bibinfo{person}{Jon McCormack}.} \bibinfo{year}{2020}\natexlab{a}.
\newblock \showarticletitle{Mirror ritual: An affective interface for emotional self-reflection}. In \bibinfo{booktitle}{\emph{Proceedings of the 2020 CHI conference on human factors in computing systems}}. \bibinfo{pages}{1--13}.
\newblock


\bibitem[Rajcic and McCormack(2020b)]%
        {rajcic2020mirror_meaning}
\bibfield{author}{\bibinfo{person}{Nina Rajcic} {and} \bibinfo{person}{Jon McCormack}.} \bibinfo{year}{2020}\natexlab{b}.
\newblock \showarticletitle{Mirror Ritual: human-machine co-construction of emotion}. In \bibinfo{booktitle}{\emph{Proceedings of the Fourteenth International Conference on Tangible, Embedded, and Embodied Interaction}}. \bibinfo{pages}{697--702}.
\newblock


\bibitem[Reed et~al\mbox{.}(2023)]%
        {reed2023negotiating}
\bibfield{author}{\bibinfo{person}{Courtney~N Reed}, \bibinfo{person}{Paul Strohmeier}, {and} \bibinfo{person}{Andrew~P McPherson}.} \bibinfo{year}{2023}\natexlab{}.
\newblock \showarticletitle{Negotiating Experience and Communicating Information Through Abstract Metaphor}. In \bibinfo{booktitle}{\emph{Proceedings of the 2023 CHI Conference on Human Factors in Computing Systems}}. \bibinfo{pages}{1--16}.
\newblock


\bibitem[Reynolds et~al\mbox{.}(2000)]%
        {reynolds2000effectiveness}
\bibfield{author}{\bibinfo{person}{Matthew~W Reynolds}, \bibinfo{person}{Laura Nabors}, {and} \bibinfo{person}{Anne Quinlan}.} \bibinfo{year}{2000}\natexlab{}.
\newblock \showarticletitle{The effectiveness of art therapy: does it work?}
\newblock \bibinfo{journal}{\emph{Art Therapy}} \bibinfo{volume}{17}, \bibinfo{number}{3} (\bibinfo{year}{2000}), \bibinfo{pages}{207--213}.
\newblock


\bibitem[Rombach et~al\mbox{.}(2022)]%
        {rombach2022high}
\bibfield{author}{\bibinfo{person}{Robin Rombach}, \bibinfo{person}{Andreas Blattmann}, \bibinfo{person}{Dominik Lorenz}, \bibinfo{person}{Patrick Esser}, {and} \bibinfo{person}{Bj{\"o}rn Ommer}.} \bibinfo{year}{2022}\natexlab{}.
\newblock \showarticletitle{High-resolution image synthesis with latent diffusion models}. In \bibinfo{booktitle}{\emph{Proceedings of the IEEE/CVF conference on computer vision and pattern recognition}}. \bibinfo{pages}{10684--10695}.
\newblock


\bibitem[Russell(2003)]%
        {russell2003core}
\bibfield{author}{\bibinfo{person}{James~A Russell}.} \bibinfo{year}{2003}\natexlab{}.
\newblock \showarticletitle{Core affect and the psychological construction of emotion.}
\newblock \bibinfo{journal}{\emph{Psychological review}} \bibinfo{volume}{110}, \bibinfo{number}{1} (\bibinfo{year}{2003}), \bibinfo{pages}{145}.
\newblock


\bibitem[Russell(2015)]%
        {russell2015greater}
\bibfield{author}{\bibinfo{person}{James~A Russell}.} \bibinfo{year}{2015}\natexlab{}.
\newblock \showarticletitle{The greater constructionist project for emotion}.
\newblock \bibinfo{journal}{\emph{The psychological construction of emotion}} (\bibinfo{year}{2015}), \bibinfo{pages}{429--447}.
\newblock


\bibitem[Saha et~al\mbox{.}(2021)]%
        {saha2021life}
\bibfield{author}{\bibinfo{person}{Koustuv Saha}, \bibinfo{person}{Jordyn Seybolt}, \bibinfo{person}{Stephen~M Mattingly}, \bibinfo{person}{Talayeh Aledavood}, \bibinfo{person}{Chaitanya Konjeti}, \bibinfo{person}{Gonzalo~J Martinez}, \bibinfo{person}{Ted Grover}, \bibinfo{person}{Gloria Mark}, {and} \bibinfo{person}{Munmun De~Choudhury}.} \bibinfo{year}{2021}\natexlab{}.
\newblock \showarticletitle{What life events are disclosed on social media, how, when, and by whom?}. In \bibinfo{booktitle}{\emph{Proceedings of the 2021 CHI conference on human factors in computing systems}}. \bibinfo{pages}{1--22}.
\newblock


\bibitem[Scarpelli et~al\mbox{.}(2019)]%
        {scarpelli2019functional}
\bibfield{author}{\bibinfo{person}{Serena Scarpelli}, \bibinfo{person}{Chiara Bartolacci}, \bibinfo{person}{Aurora D'Atri}, \bibinfo{person}{Maurizio Gorgoni}, {and} \bibinfo{person}{Luigi De~Gennaro}.} \bibinfo{year}{2019}\natexlab{}.
\newblock \showarticletitle{The functional role of dreaming in emotional processes}.
\newblock \bibinfo{journal}{\emph{Frontiers in psychology}}  \bibinfo{volume}{10} (\bibinfo{year}{2019}), \bibinfo{pages}{459}.
\newblock


\bibitem[Scott et~al\mbox{.}(2023)]%
        {scott2023trauma}
\bibfield{author}{\bibinfo{person}{Carol~F Scott}, \bibinfo{person}{Gabriela Marcu}, \bibinfo{person}{Riana~Elyse Anderson}, \bibinfo{person}{Mark~W Newman}, {and} \bibinfo{person}{Sarita Schoenebeck}.} \bibinfo{year}{2023}\natexlab{}.
\newblock \showarticletitle{Trauma-Informed Social Media: Towards Solutions for Reducing and Healing Online Harm}. In \bibinfo{booktitle}{\emph{Proceedings of the 2023 CHI Conference on Human Factors in Computing Systems}}. \bibinfo{pages}{1--20}.
\newblock


\bibitem[Slayton et~al\mbox{.}(2010)]%
        {slayton2010outcome}
\bibfield{author}{\bibinfo{person}{Sarah~C Slayton}, \bibinfo{person}{Jeanne D'Archer}, {and} \bibinfo{person}{Frances Kaplan}.} \bibinfo{year}{2010}\natexlab{}.
\newblock \showarticletitle{Outcome studies on the efficacy of art therapy: A review of findings}.
\newblock \bibinfo{journal}{\emph{Art therapy}} \bibinfo{volume}{27}, \bibinfo{number}{3} (\bibinfo{year}{2010}), \bibinfo{pages}{108--118}.
\newblock


\bibitem[Smith(2007)]%
        {smith2007hermeneutics}
\bibfield{author}{\bibinfo{person}{Jonathan~A Smith}.} \bibinfo{year}{2007}\natexlab{}.
\newblock \showarticletitle{Hermeneutics, human sciences and health: Linking theory and practice}.
\newblock \bibinfo{journal}{\emph{International Journal of Qualitative Studies on health and Well-being}} \bibinfo{volume}{2}, \bibinfo{number}{1} (\bibinfo{year}{2007}), \bibinfo{pages}{3--11}.
\newblock


\bibitem[Stilwell et~al\mbox{.}(2021)]%
        {stilwell2021painful}
\bibfield{author}{\bibinfo{person}{Peter Stilwell}, \bibinfo{person}{Christie Stilwell}, \bibinfo{person}{Brenda Sabo}, {and} \bibinfo{person}{Katherine Harman}.} \bibinfo{year}{2021}\natexlab{}.
\newblock \showarticletitle{Painful metaphors: enactivism and art in qualitative research}.
\newblock \bibinfo{journal}{\emph{Medical humanities}} \bibinfo{volume}{47}, \bibinfo{number}{2} (\bibinfo{year}{2021}), \bibinfo{pages}{235--247}.
\newblock


\bibitem[Stuckey and Nobel(2010)]%
        {stuckey2010connection}
\bibfield{author}{\bibinfo{person}{Heather~L Stuckey} {and} \bibinfo{person}{Jeremy Nobel}.} \bibinfo{year}{2010}\natexlab{}.
\newblock \showarticletitle{The connection between art, healing, and public health: A review of current literature}.
\newblock \bibinfo{journal}{\emph{American journal of public health}} \bibinfo{volume}{100}, \bibinfo{number}{2} (\bibinfo{year}{2010}), \bibinfo{pages}{254--263}.
\newblock


\bibitem[Terzimehi{\'c} et~al\mbox{.}(2021)]%
        {terzimehic2021memeories}
\bibfield{author}{\bibinfo{person}{Na{\dj}a Terzimehi{\'c}}, \bibinfo{person}{Svenja~Yvonne Sch{\"o}tt}, \bibinfo{person}{Florian Bemmann}, {and} \bibinfo{person}{Daniel Buschek}.} \bibinfo{year}{2021}\natexlab{}.
\newblock \showarticletitle{MEMEories: internet memes as means for daily journaling}. In \bibinfo{booktitle}{\emph{Designing Interactive Systems Conference 2021}}. \bibinfo{pages}{538--548}.
\newblock


\bibitem[Tholander and Normark(2020)]%
        {tholander2020crafting}
\bibfield{author}{\bibinfo{person}{Jakob Tholander} {and} \bibinfo{person}{Maria Normark}.} \bibinfo{year}{2020}\natexlab{}.
\newblock \showarticletitle{Crafting personal information-resistance, imperfection, and self-creation in bullet journaling}. In \bibinfo{booktitle}{\emph{Proceedings of the 2020 CHI Conference on Human Factors in Computing Systems}}. \bibinfo{pages}{1--13}.
\newblock


\bibitem[Torre and Lieberman(2018)]%
        {torre2018putting}
\bibfield{author}{\bibinfo{person}{Jared~B Torre} {and} \bibinfo{person}{Matthew~D Lieberman}.} \bibinfo{year}{2018}\natexlab{}.
\newblock \showarticletitle{Putting feelings into words: Affect labeling as implicit emotion regulation}.
\newblock \bibinfo{journal}{\emph{Emotion Review}} \bibinfo{volume}{10}, \bibinfo{number}{2} (\bibinfo{year}{2018}), \bibinfo{pages}{116--124}.
\newblock


\bibitem[Wilson and Schooler(1991)]%
        {wilson1991thinking}
\bibfield{author}{\bibinfo{person}{Timothy~D Wilson} {and} \bibinfo{person}{Jonathan~W Schooler}.} \bibinfo{year}{1991}\natexlab{}.
\newblock \showarticletitle{Thinking too much: introspection can reduce the quality of preferences and decisions.}
\newblock \bibinfo{journal}{\emph{Journal of personality and social psychology}} \bibinfo{volume}{60}, \bibinfo{number}{2} (\bibinfo{year}{1991}), \bibinfo{pages}{181}.
\newblock


\bibitem[Wright et~al\mbox{.}(2008)]%
        {wright2008aesthetics}
\bibfield{author}{\bibinfo{person}{Peter Wright}, \bibinfo{person}{Jayne Wallace}, {and} \bibinfo{person}{John McCarthy}.} \bibinfo{year}{2008}\natexlab{}.
\newblock \showarticletitle{Aesthetics and experience-centered design}.
\newblock \bibinfo{journal}{\emph{ACM Transactions on Computer-Human Interaction (TOCHI)}} \bibinfo{volume}{15}, \bibinfo{number}{4} (\bibinfo{year}{2008}), \bibinfo{pages}{1--21}.
\newblock


\bibitem[Zarei et~al\mbox{.}(2020)]%
        {zarei2020investigating}
\bibfield{author}{\bibinfo{person}{Niloofar Zarei}, \bibinfo{person}{Sharon~Lynn Chu}, \bibinfo{person}{Francis Quek}, \bibinfo{person}{Nanjie'Jimmy' Rao}, {and} \bibinfo{person}{Sarah~Anne Brown}.} \bibinfo{year}{2020}\natexlab{}.
\newblock \showarticletitle{Investigating the Effects of Self-Avatars and Story-Relevant Avatars on Children's Creative Storytelling}. In \bibinfo{booktitle}{\emph{Proceedings of the 2020 CHI Conference on Human Factors in Computing Systems}}. \bibinfo{pages}{1--11}.
\newblock


\bibitem[Zhang et~al\mbox{.}(2022)]%
        {zhang2022storydrawer}
\bibfield{author}{\bibinfo{person}{Chao Zhang}, \bibinfo{person}{Cheng Yao}, \bibinfo{person}{Jiayi Wu}, \bibinfo{person}{Weijia Lin}, \bibinfo{person}{Lijuan Liu}, \bibinfo{person}{Ge Yan}, {and} \bibinfo{person}{Fangtian Ying}.} \bibinfo{year}{2022}\natexlab{}.
\newblock \showarticletitle{StoryDrawer: A Child--AI Collaborative Drawing System to Support Children's Creative Visual Storytelling}. In \bibinfo{booktitle}{\emph{Proceedings of the 2022 CHI Conference on Human Factors in Computing Systems}}. \bibinfo{pages}{1--15}.
\newblock


\end{thebibliography}
